\documentclass[%
reprint,
superscriptaddress,
showpacs,
amsmath,amssymb,
aps,
]{revtex4-2}
\usepackage{graphicx}% Include figure files
\usepackage{dcolumn}% Align table columns on decimal point
\usepackage{bm}% bold math
\usepackage{CJKutf8}
\usepackage{color}
\usepackage{amssymb}
\usepackage{amsfonts}
\usepackage{palatino}
\usepackage[colorlinks,urlcolor=blue,linkcolor=blue,citecolor=blue,anchorcolor=blue]{hyperref}
\makeatletter

\newcommand{\Rmnum}[1]{\expandafter\@slowromancap\romannumeral #1@}

\newcommand{\bk}{{\bm k}}
\newcommand{\bK}{{\bf K}}

\makeatother
\begin{document}

\begin{CJK*}{UTF8}{gbsn}

\preprint{APS/123-QED}

\title{Bistable topological edge states in polariton microcavities with unpaired Dirac cones}

\author{Zhuo Zhang}
\affiliation{Key Laboratory for Physical Electronics and Devices, Ministry of Education, School of Electronic Science and Engineering, Xi'an Jiaotong University, Xi'an 710049, China}

\author{Yaroslav V. Kartashov}
\affiliation{Institute of Spectroscopy, Russian Academy of Sciences, Troitsk, Moscow, 108840, Russia} 

\author{Yongdong Li}
\affiliation{Key Laboratory for Physical Electronics and Devices, Ministry of Education, School of Electronic Science and Engineering, Xi'an Jiaotong University, Xi'an 710049, China}

\author{Zhen-Nan Tian}
\author{Qi-Dai Chen}
\affiliation{State Key Laboratory of Integrated Optoelectronics, College of Electronic Science and Engineering, Jilin University, Changchun 130012, China}

\author{Yiqi Zhang}
\email{zhangyiqi@xjtu.edu.cn}
\affiliation{Key Laboratory for Physical Electronics and Devices, Ministry of Education, School of Electronic Science and Engineering, Xi'an Jiaotong University, Xi'an 710049, China}

\date{\today}% It is always \today, today,

\begin{abstract}
\noindent
Among the most intriguing properties of honeycomb lattices is the presence of Dirac points that typically emerge in pairs, which can be destroyed by physical effects breaking certain symmetries of the system and leading to nontrivial band topology. We propose a nonlinear microcavity system supporting condensate of exciton-polaritons, where simultaneous breakup of inversion and time-reversal symmetries results in unusual spectrum with \textit{unpaired} Dirac cones, profoundly affecting the properties of unidirectional edge states. Realized as an array of microcavity pillars, the inversion symmetry is broken by \textit{fission} of pillar belonging to one of sublattices of honeycomb array into three pillars, while time-reversal symmetry is broken due to interplay of Zeeman splitting in the external magnetic field and spin-orbit coupling. Despite the absence of complete spectral gap, unidirectional edge states may still emerge that can circumvent array corners. Resonant optical pumping leads to reach bistability effects and allow selective excitation of the edge states. We obtain first example of stable \textit{localized dissipative} edge soliton that circulates along the periphery of insulator over indefinitely long times without radiation. Our results suggest a new platform for nonlinear topological photonics and reveal nontrivial interplay between unpaired Dirac cones and nonlinear effects.
\end{abstract}

\maketitle

\end{CJK*}

\section{Introduction}

The presence of specific degeneracies in linear spectra of certain classes of lattices, such as Dirac cones in spectra of honeycomb and kagome lattices, is one of the major prerequisites for realization of nontrivial topological phases in such structures. Dirac cones in spectra of such lattices usually emerge in pairs in non-equivalent $\textbf{K}$ and $\textbf{K}'$ points of the Brillouin zone. Such lattices are paradigmatic platforms for the implementation of topological phases under the action of physical effects breaking certain symmetries of the system, leading to simultaneous opening of all Dirac cones and the appearance of the complete topological spectral gap. However, a particularly intriguing situation emerges when \textit{unpaired} Dirac cones appear in the spectrum of the system (i.e. when Dirac cones persist, say, in $\textbf{K}'$ points, but are opened in $\textbf{K}$ points of the Brillouin zone), as was predicted in the frames of the Haldane model~\cite{haldane.prl.61.2015.1988} and experimentally demonstrated in ultracold atoms~\cite{jotzu.nature.515.237.2014}.

The investigation of the fate of the topological edge states in such unusual situation without complete topological spectral gap is of considerable interest for different areas of physics and is still in its infancy. Thus, in photonic topological systems~\cite{lu.np.8.821.2014, ozawa.rmp.91.015006.2019, zhang.nature.618.687.2023, leykam.nrp.8.55.2026} the settings, where the unpaired Dirac cones could emerge, remain rather scarce. In particular, in topological waveguide arrays the waveguide helicity (leading to effective gauge field that breaks time-reversal symmetry~\cite{rechtsman.nature.496.196.2013}) for asynchronous waveguide rotations (breaking also inversion symmetry~\cite{leykam.prl.117.013902.2016} of the system) may lead to the emergence of the unpaired Dirac cones. A different approach to realization of the unpaired Dirac cones relies on the introduction of detuning in two sublattices in helical honeycomb waveguide arrays~\cite{zhong.pr.12.2078.2024} thereby combining physics of valley-Hall~\cite{xiao.prl.99.236809.2007, noh.prl.120.063902.2018, xue.apr.2.2100013.2021, tang.lpr.1.2100300.2022} and Floquet topological systems. Other approaches to the realization of the unpaired Dirac cones in magneto-optic or non-Hermitian systems~\cite{zhou.prb.98.205115.2018,xue.prl.124.236403.2020,wang.nc.14.4457.2023} have been suggested, but, to the best of our knowledge, the only experimental demonstration of such unpaired cones has been reported in gyromagnetic photonic crystals in microwave range~\cite{liu.nc.1873.2020}.

In this work we propose a new setting, where unpaired Dirac cones can be observed, and where associated unusual edge states can be registered. Our proposal relies on essentially \textit{dissipative and nonlinear} polariton microcavity platform. Etching of such microcavities allows creation of a honeycomb or kagome array of pillars, where condensation of polaritons is possible in the presence of non-resonant or resonant optical pumping. Microcavities supporting polariton condensates represent a well-established platform for realization of topological insulators of various types, including Chern~\cite{klembt.nature.562.552.2018, nalitov.prl.114.116401.2015, bardyn.prb.91.161413.2015, karzig.prx.5.031001.2015, milivic.2d.2.034012.2015, gulevich.prb.94.115437.2016, zhang.lpr.12.1700348.2018, zhang.apl.3.120801.2018, zhang.lpr.13.1900198.2019}, valley-Hall~\cite{liu.science.370.600.2020, peng.nn.2024, jin.nc.15.10563.2024}, higher-order~\cite{wu.sa.9.eadg4322.2023, su.sa.7.8049.2021, bennenhei.acs.11.3046.2024, jin.nc.16.6002.2025} insulators and their one-dimensional counterparts~\cite{jean.np.11.651.2017}. It has been also demonstrated that Floquet engineering~\cite{redondo.np.18.548.2024} and non-Hermitian topology~\cite{liang.np.22.151.2026} can be realized in polariton condensates. Recent progress in the investigation of topological states in polaritonic systems is described in reviews~\cite{solnyshkov.ome.11.1119.2021, luo.apr.10.011316.2023}. However, in all polariton Chern insulators constructed so far on honeycomb or kagome arrays Dirac cones appear in pairs and are simultaneously destroyed upon breakup of time-reversal symmetry of the system.

To realize polariton system with unpaired Dirac cones in spectrum, here we consider \textit{fissioned} honeycomb array of microcavity pillars, where inside each unit cell a pillar belonging to one of sublattices is replaced by three close pillars. This transformation breaks inversion symmetry of the system and leads to opening of the Dirac cones in spectrum, but subsequent breakup also of time-reversal symmetry due to combined action of Zeeman splitting in the external magnetic field and spin-orbit coupling restores the cones in $\textbf{K}'$ points of the Brillouin zone, while increasing the gap around $\textbf{K}$ points. We show that despite the fact that this system does not feature a complete topological gap, unidirectional edge states still appear in its spectrum upon array truncation, and that they demonstrate signatures of topological protection by bypassing corners of the structure and nested defects without backscattering or radiation into bulk. The proposed system is dissipative and strongly nonlinear (this was previously employed for demonstration of various nonlinear phenomena with topological polaritons, such as modulation instability and soliton formation~\cite{kartashov.optica.3.1228.2016, gulevich.sr.7.1780.2017, bleu.prb.93.085438.2016, kartashov.prl.119.253904.2017, li.prb.97.081103.2018, bleu.nc.9.3991.2018, kartashov.prl.122.083902.2019, zhang.pra.99.053836.2019}, see also reviews~\cite{smirnova.apr.7.021306.2020, szameit.np.20.905.2024}) and requires external optical pumping. We show that selective excitation of topological edge states in system with unpaired Dirac cones can be achieved with resonant optical pumping. For the first time to our knowledge, we obtain localized topological edge soliton that circulates under resonant pumping along the periphery of the microcavity without any signs of decay or radiation, i.e. existing as long as pumping is on. This contribution may lead to further development of theory of topological edge solitons~\cite{lumer.prl.111.243905.2013, leykam.prl.117.143901.2016, zhang.prl.123.254103.2019, ivanov.acs.7.735.2020, mukherjee.science.368.856.2020, maczewsky.science.370.701.2020, mukherjee.prx.11.041057.2021, zhong.ap.3.056001.2021, tang.oe.29.39755.2021, ren.nano.10.3559.2021, tang.rrp.74.504.2022} to dissipative nonlinear systems.

\section{Unpaired Dirac cones in structured polariton microcavity}

We consider here the evolution of spinor wavefunction ${\mathbf{\Psi} = (\psi_+, \psi_-)^{\rm T}}$ written in circular polarization basis $\psi_\pm=(\psi_x\mp\psi_y)/2^{1/2}$ and describing linearly and nonlinearly coupled spin-positive $\psi_+$ and spin-negative $\psi_-$ components of polariton condensate in the array of microcavity pillars, which can be fabricated using microcavity etching~\cite{solnyshkov.ome.11.1119.2021}. The evolution of the wavefunction components is described by the coupled dimensionless Gross-Pitaevskii equations:
\begin{align}\label{eq1}
	i \frac{\partial \psi_{\pm}}{\partial t} = & -\frac{1}{2} \left( \frac{\partial^2}{\partial x^2} + \frac{\partial^2}{\partial y^2} \right) \psi_{\pm} 
	+\beta \left( \frac{\partial}{\partial x} \mp i \frac{\partial}{\partial y} \right)^2 \psi_{\mp} \notag \\
	& + [ \mathcal{R}(x,y) \pm \Omega  - i\gamma ] \psi_{\pm} \notag \\
	& + (|\psi_{\pm}|^2 + \sigma|\psi_{\mp}|^2)\psi_{\pm} + \mathcal{H}_\pm(x,y,t),
\end{align}
where the first term on the right-hand side describes dispersion of polaritons near the bottom of the lower polariton branch; the term $\sim \beta$  term describes spin-orbit coupling stemming from TE-TM splitting of microcavity modes~\cite{sala.prx.5.011034.2015, dufferwiel.prl.115.246401.2015}; the term $\sim \Omega$ stands for the magnitude of the Zeeman splitting induced by the external uniform magnetic field applied to the microcavity~\cite{klembt.nature.562.552.2018}; $\gamma$ is the coefficient of linear losses that we assume identical in $\psi_\pm$ components; we also take into account repulsive interactions of polaritons with the same spin, but weakly attractive interactions between polaritons with opposite spins, whose strength is determined by the parameter $\sigma$; while ${\mathcal{H}_\pm(x,y,t)=h_\pm (x,y)e^{-i\epsilon t}}$ stands for resonant optical pump that can be spatially inhomogeneous, $\epsilon$ is the pump frequency detuning from the bottom of the lower polariton branch. The function ${\mathcal{R}(x,y) = -p \sum_{m,n} e^{ -(\bm{r}-\bm{r}_{m,n})^2/\delta^2}}$ approximates the potential energy landscape in microcavity and contains contributions from all microcavity pillars described by Gaussian functions with characteristic width $\delta$, depth $p$, and coordinates ${\bm{r}_{m,n}=(x_{m,n},y_{m,n})}$. 

\begin{figure}[h!]
	\centering
	\includegraphics[width=\columnwidth]{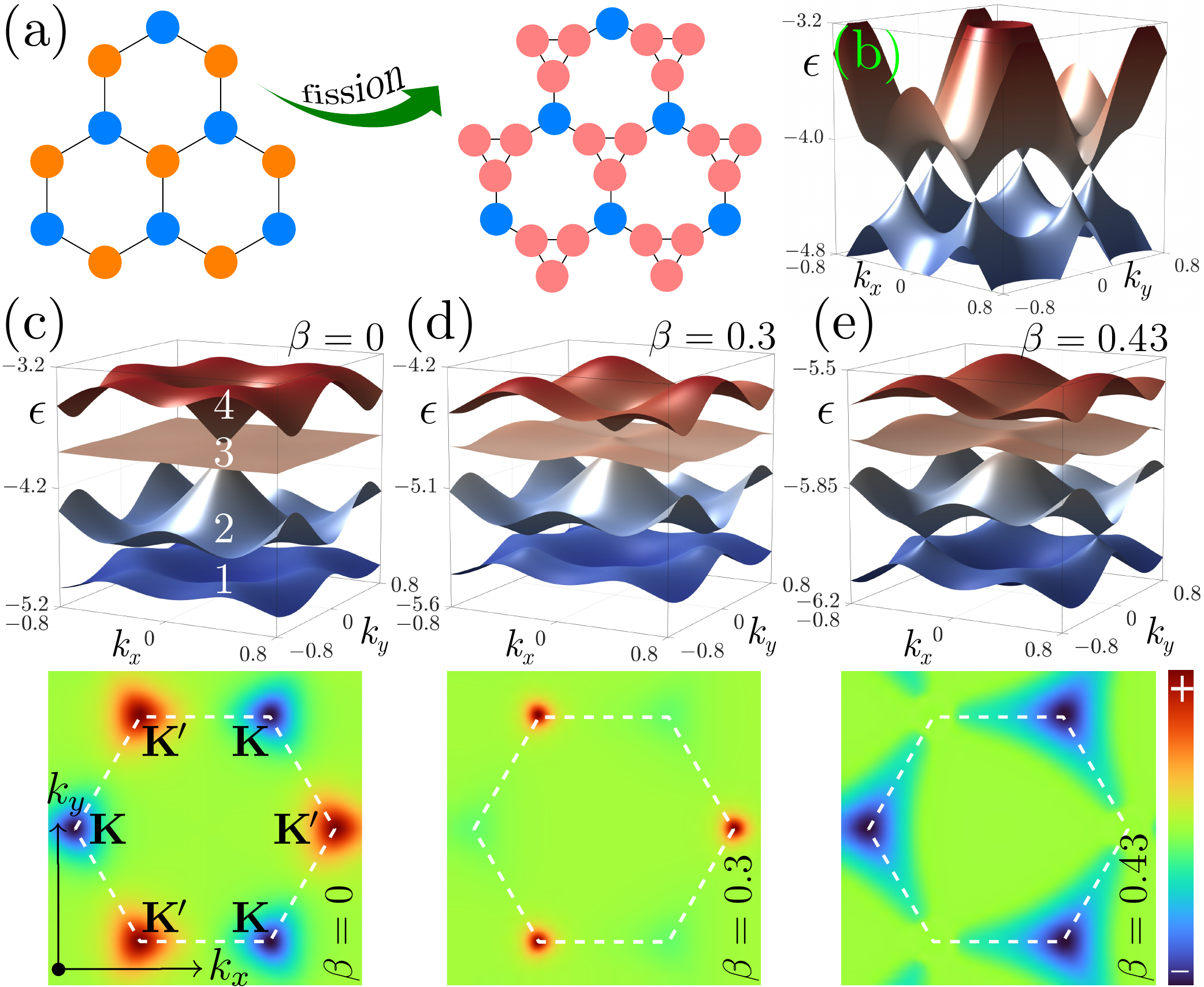}
	\caption{%\textbf{Fissioned honeycomb array and its band structure.}
		(a) Honeycomb array with two sublattices indicated by orange and blue dots and its fissioned version, where pillars corresponding to orange dots have been transformed into three pink pillars. Spacing in fissioned array is the same for all pillars and is given by $d$. (b) Band structure of the honeycomb array at $\beta=0$. (c-e) Top row shows band structures of the fissioned honeycomb arrays for different spin-orbit coupling strengths $\beta$, while bottom row shows corresponding Berry curvature for band $1$. Unpaired Dirac cones appear in panel (e) at ${\beta\approx 0.43}$. White dashed hexagon shows the first Brillouin zone with coordinates of the $\bf K$ points being ${(1/3,\pm1/3^{1/2}){\rm K}_x}$ and ${(-2/3,0){\rm K}_x}$, while those of ${\bf K}'$ points being ${(-1/3,\pm1/3^{1/2}){\rm K}_x}$ and ${(2/3,0){\rm K}_x}$, where ${{\rm K}_x=2\pi/{\rm X}}$ and ${{\rm X}=(3^{1/2}+1)d}$.}
	\label{fig1}
\end{figure}

In Eq.~(\ref{eq1}) we normalized $x,y$ coordinates to the characteristic spatial scale ${r_0\sim1~\mu \rm{m}}$; evolution time is normalized to $\hbar \varepsilon_0^{-1}$, where ${\varepsilon_0=\hbar^2/mr_0^2}$ is the characteristic energy; ${m=2m_xm_y/(m_x+m_y)}$ is the effective polariton mass with $m_{x,y}$ being the effective masses of the TM (subscript $x$) and TE (subscript $y$) polaritons, respectively; spin-orbit coupling strength is expressed as ${2\beta=(m_x-m_y)/(m_x+m_y)}$. All energy parameters (including the depth of potential wells and Zeeman splitting) in Eq. (\ref{eq1}) are normalized to the characteristic energy $\varepsilon_0$. For details of normalizations, see~\cite{kartashov.optica.3.1228.2016, zhang.lpr.12.1700348.2018}. Further we adopt the dimensionless values of parameters ${p=10}$, ${\delta=0.5}$, ${\sigma=-0.05}$, ${\Omega=0.9}$, and ${d=1.4}$ (spacing between neighboring micropillars in fissioned array).

We are interested in transformation of the linear spectrum of this system under the action of fission applied to original honeycomb array of microcavity pillars, Zeeman splitting and spin-orbit coupling and consider first unconstrained, truly periodic array schematically depicted in Fig. \ref{fig1}(a). We preliminary omit dissipative effects considering the system in the absence of resonant pump and losses ${(h_\pm=0, \gamma=0)}$ and polariton-polariton interactions. Linear eigenmodes of bulk array are Bloch waves ${\psi_{\pm}(x,y,t)=u_{\pm}(x,y) \exp(ik_x x + i k_y y-i\epsilon t)}$, where ${u_{\pm}(x,y)}$ is the spatially periodic part of the Bloch wavefunction with energy $\epsilon$, and $k_{x,y}$ are the two components of the Bloch momentum $\bm{k}$. Corresponding Bloch waves and their energies can be calculated with the above substitution from Eq.~(\ref{eq1}) using plane-wave expansion method.

Standard honeycomb micropillar array consists of two sublattices, as indicated by blue and orange circles in Fig.~\ref{fig1}(a). Spacing between neighboring pillars in such array is set as ${d'=(1+3^{-1/2})d}$. At $\beta=0$ the spectrum of such array contains two identical groups of bands shifted along the $\epsilon$ axis by $2\Omega$, with six Dirac cones clearly visible between two lowest bands in each group (further we consider only lowest bands characterized for our parameters by the dominance of $\psi_-$ component over $\psi_+$ one), as shown in linear spectrum $\epsilon(k_x,k_y)$ in Fig.~\ref{fig1}(b). Fission of pillars belonging to orange sublattice into three pillars shown pink in Fig.~\ref{fig1}(a) (they reside in the corners of triangle, whose center coincides with center of former orange pillar, while spacing between neighboring sites in the entire array after fission becomes equal to $d$) substantially enriches the spectrum. Its lowest four bands at $\beta=0$ are shown in Fig.~\ref{fig1}(c). One can see that ``fermionic'' Dirac cones between bands $1$ and $2$ are destroyed by fission procedure that results in the appearance of the complete spectral gap. Instead, one can observe the emergence of a ``bosonic'' Dirac cone~\cite{leykam.aipx.1.101.2016} between bands $2$ and $4$ intersected by nearly flat band $3$.

% Berry curvature of the band with index $n$ can be calculated as 
% \begin{align}
	% {\mathcal B}_n(\bm{k})=i \sum_{n' \neq n}  \left( \frac{\langle v_n|\partial_{k_x} H| v_{n'}\rangle \langle v_{n'}|\partial_{k_y} H| v_n\rangle}{(\epsilon_n-\epsilon_{n'})^2} - 
	% \frac{\langle v_n|\partial_{k_y} H| v_{n'}\rangle \langle v_{n'}|\partial_{k_x} H| v_n\rangle}{(\epsilon_n-\epsilon_{n'})^2}\right),
	% \end{align}
% where $|v_n \rangle$ and $\epsilon_n$ are the \yz{Bloch functions (i.e., eigenvectors)} and their energies for \yz{the linear Hamiltonian of the system $H$ by neglecting the nonlinear term, the pumping and the loss in Eq.~(\ref{eq1})}, and $\langle \cdot | \cdot \rangle$ is the inner product.

Remarkably, increasing strength of spin-orbit coupling $\beta$ leads to gradual restoration of ``fermionic" Dirac cones between bands $1$ and $2$ around $\textbf{K}'$ points of the Brillouin zone, while causing further splitting of these bands around $\textbf{K}$ points, as shown in Figs.~\ref{fig1}(d) and \ref{fig1}(e). As a result, three \textit{unpaired} Dirac cones emerge in spectrum around $\beta\approx0.43$, as shown in Fig.~\ref{fig1}(e). ``Bosonic" cone is instead destroyed with increase of $\beta$. This transformation of spectrum is accompanied by qualitative modifications in Berry curvature ${\mathcal B}_n(\bm{k})$, where $n$ is the band index~\cite{xiao.rmp.82.1959.2010, fuchs.epjb.77.351.2010, fukui.jpsj.74.1674.2005}. Representative $\mathcal{B}_n(\bm{k})$ distributions for band ${n=1}$ are shown in the bottom row of Fig.~\ref{fig1} for various amplitudes of spin-orbit coupling $\beta$. These distributions are tightly connected with topological properties of the bulk bands of this system characterized by the Chern numbers $C_n$.
% \begin{equation}
	% { C}_{n}=\frac{1}{2 \pi} \int_{\rm BZ}  {\mathcal B}_{n}(\bm{k}) d \bm{k},
	% \end{equation}
% where $\rm BZ$ indicates that the integration is carried over the first Brillouin zone. 

Thus, at ${\beta=0}$ the Berry curvature around $\bf K$ and $\bf K'$ points (in corresponding valleys) has opposite sign, but equal magnitude, yielding Chern number $C_1=0$, which means that the complete gap that lies above band $1$ is topologically trivial. For intermediate values of ${\beta \sim 0.3}$ the gap between two bands at $\bf K'$ points is substantially smaller than that at $\bf K$ points, leading to strong positive peaks in $\mathcal{B}_1(\bm{k})$ around $\textbf{K}'$ points, while around $\textbf{K}$ the curvature remains negative and broad. Still the integration in the entire $\rm BZ$ yields ${C_1=0}$. The situation changes qualitatively when \textit{unpaired} Dirac cones appear in $\textbf{K}'$ points at ${\beta \approx 0.43}$. In this case Berry curvature is not defined in $\bf K'$ points, as two bands touch each other.
%while integration of $\mathcal{B}_1(\bm{k})$ around $\bf K$ valleys only yields ${C_1=-0.5}$.
We are interested in properties of the edge states in this critical case with \textit{unpaired} Dirac cones, where topology of the bands changes qualitatively. Indeed, further increase of $\beta$ leads to reopening of topological gap between two bands with Chern numbers ${C_1=-1}$ and ${C_2=+1}$, i.e. the transition to classical Chern insulator phase.
However, the Chern insulator phase exists in a very narrow range of $\beta$.

\section{Linear topological edge states}

\begin{figure*}[htbp]
	\centering
	\includegraphics[width=\textwidth]{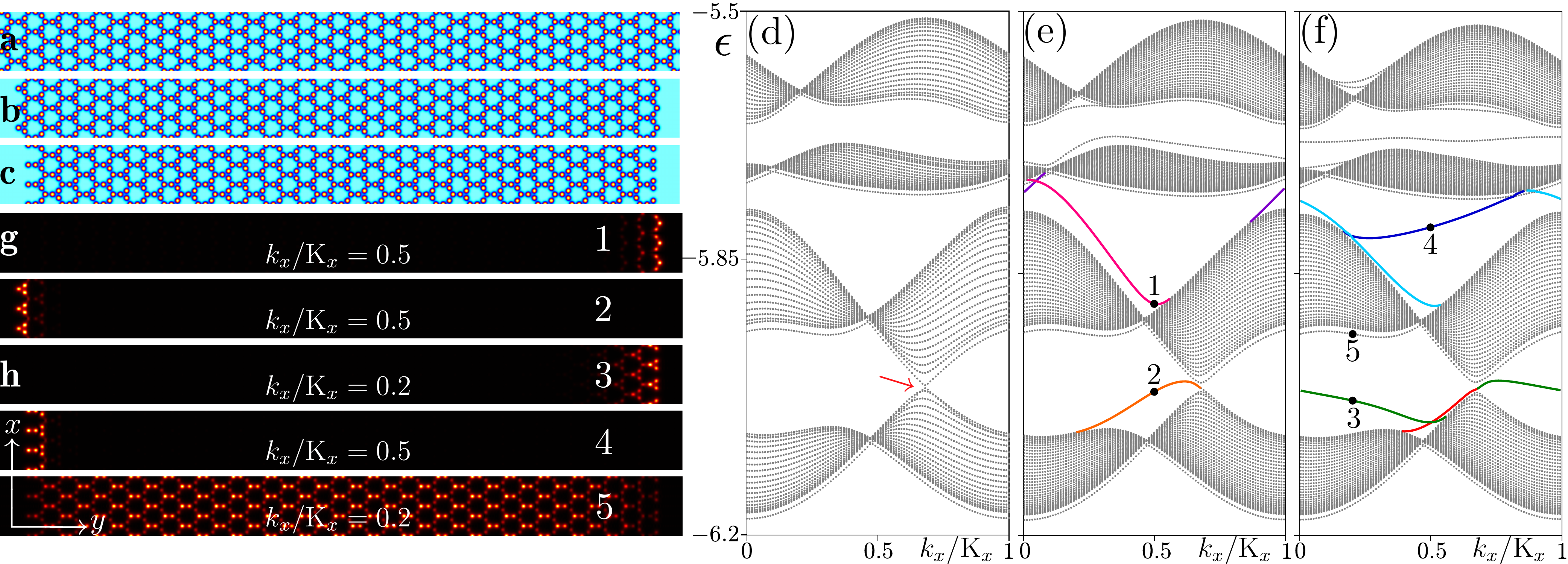}
	\caption{%\textbf{Edge states in the fissioned honeycomb lattice.}
		(a) Fissioned array with integer number of cells on finite $y$-window (here $y$ axis is horizontal) and its ``projected" linear spectrum $\epsilon(k_x)$ in (d), where unpaired Dirac cone is indicated by the red arrow. (b) Finite ribbon with zigzag-zigzag boundaries in $y$ made from fissioned array, and its linear spectrum in (e). (c) Finite ribbon with bearded-bearded boundaries in $y$ and its linear spectrum in (f). Edge states in (e) and (f) are shown with colored curves. (g) Density distributions in the edge states (only dominating $|\psi_-|^2$ component is shown) corresponding to the dots $1$ and $2$ in (e). (h) Density distributions in the edge states corresponding to the dots ${3\sim5}$ in (f). Distributions and arrays in the left column are shown within the window ${-30d'\le y \le 30d'}$ and ${-1.5{\rm X}\le x \le 1.5{\rm X}}$. In all cases ${\beta=0.43}$ and ${\Omega=0.9}$.}
	\label{fig2}
\end{figure*}

To understand whether edge states can exist in this system with unpaired Dirac cones, we consider a ribbon finite in the $y$-direction made from fissioned array of microcavity pillars. Because the ribbon is still periodic in $x$, its eigenmodes are Bloch waves ${\psi_{\pm}(x,y,t)=u_{\pm}(x,y) \exp(ik_x x -i\epsilon t)}$ depending on the momentum $k_x$. First we consider the array with integer number of cells on finite $y$-window and assume periodicity of the wavefunction $\psi_\pm$ on the entire $y$-window, as shown in Fig. \ref{fig2}(a) (for illustration purposes $y$ axis is made horizontal in this figure). The period of this structure in $x$ is ${\rm X}=3^{1/2}d'=(3^{1/2}+1)d$, defining the width of the Brillouin zone ${\rm K}_x=2\pi/\rm{X}$. ``Projected" spectrum $\epsilon(k_x)$ of such array at critical value of ${\beta =0.43}$ is shown in Fig.~\ref{fig2}(d), where edge states are absent, since there are no boundaries, while an unpaired Dirac cone is clearly visible, as indicated by the red arrow.

Remarkably, truncation of the array leading to zigzag-zigzag boundaries in $y$ [Fig.~\ref{fig2}(b)] results in the appearance of the edge state [orange curve in corresponding projected spectrum in Fig.~\ref{fig2}(e)] above band $1$ despite the \textit{absence} of the complete gap. Edge states also appear above band $2$, inside complete gap. Their representative profiles corresponding to dots $1,2$ are shown in Fig.~\ref{fig2}(g). These states correspond to branches with opposite slopes $\partial \epsilon/\partial k_x$; they are localized at the opposite array edges and have notably different density distributions. The array with bearded-bearded boundaries [Fig.~\ref{fig2}(c)] also admits edge states with different internal structure above band $1$ (green branch) and above band $2$ (blue branch), see projected spectrum in Fig.~\ref{fig2}(f). Examples of such states corresponding to dots $3,4$ are shown in Fig.~\ref{fig2}(h) along with bulk state corresponding to dot $5$. All states above band $1$ emanate from Dirac cone at $2\textrm{K}_x/3$. When they reside inside the gap, they may be strongly localized, but delocalize when they enter band $1$.

\begin{figure*}[t!]
	\centering
	\includegraphics[width=\textwidth]{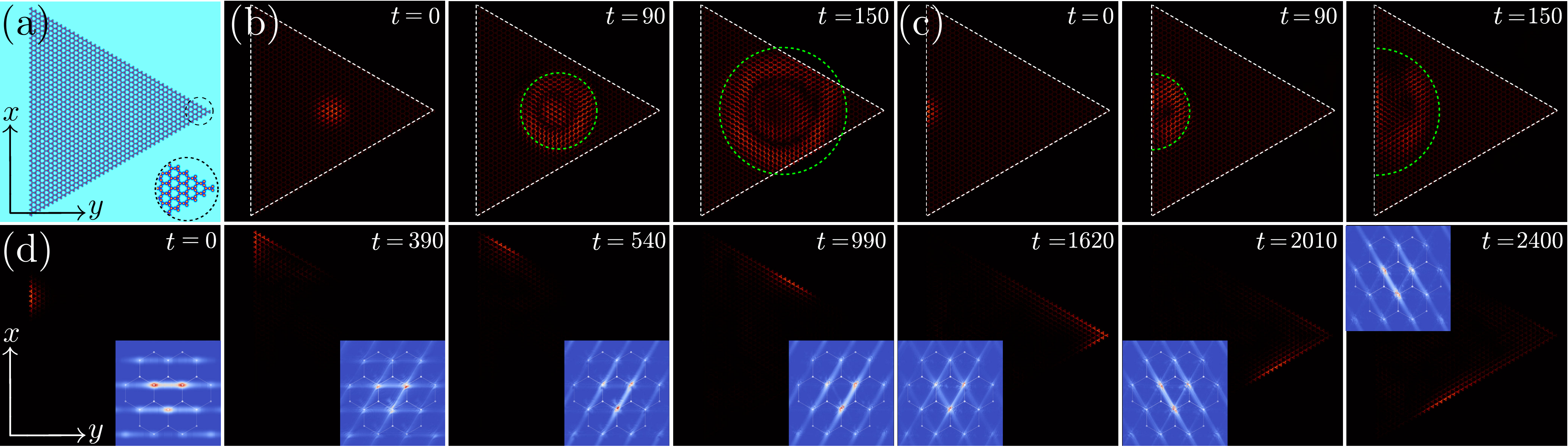}
	\caption{(a) Fissioned honeycomb array in a triangular configuration with zigzag edges. Inset shows the magnified sector of the structure.
		(b) Excitation in the bulk of triangular array corresponding to $\textbf{K}'$ unpaired Dirac cone that results in conical diffraction. Density distributions $|\psi_-|^2$ in different moments of time are shown. Green dotted ring shows theoretical prediction for the radius of conical diffraction pattern. (c) Excitation at the edge of triangular array corresponding to the $\textbf{K}'$ cone. (d) Evolution of the edge state corresponding to the momentum $\textrm{K}_x/3$ along the edge. The insets show the density distribution in Fourier space with the hexagons representing Brillouin zones. All distributions are shown within ${-86\le x,y \le 86}$ regions. White dashed triangle shows array boundaries. In all cases ${\beta=0.43}$ and ${\Omega=0.9}$.}
	\label{fig3}
\end{figure*}

Since edge states can coexist in this system with such unusual spectral features as unpaired Dirac cones, one can observe drastic differences in propagation dynamics for excitations with representative Bloch momenta $k_x=\textrm{K}_x/3$ and $k_x=2\textrm{K}_x/3$ in projected spectrum (corresponding to an in-gap edge state and the vicinity of an unpaired Dirac cone, respectively). To illustrate this, we designed a triangular array with zigzag edges shown in Fig.~\ref{fig3}(a). We used a linear bulk mode corresponding to the unpaired Dirac cone at $\textbf{K}'$ point ($k_x=2\textrm{K}_x/3$) modulated with a broad Gaussian envelope as an initial state. When this initial state is located in the bulk of the array, one observes pronounced conical diffraction, whose characteristic feature is the appearance of expanding ring of nearly constant width [Fig.~\ref{fig3}(b)]. When the same state is located at the edge of the array (we drop a part of it falling outside the array), one also observes conical diffraction into the depth of array, strongly modified by the array edge (dotted white line) [Fig.~\ref{fig3}(c)] without any displacement along the edge of the array. The radius of the ring agrees well with theoretical prediction shown with green dotted line, according to which the radius $|t\epsilon'_{\bk\simeq\bK'}|$ is determined by the slope of the band in the vicinity of the Dirac cone. In sharp contrast, if the edge state with momentum $\textrm{K}_x/3$ along the edge (hence it is localized across the edge) is used as an initial state (here we placed it at the same left array edge and imposed broad Gaussian envelope on it along the edge), it moves along the edge and circumvents array corners without backscattering and inter-valley scattering [Fig.~\ref{fig3}(d)] thereby showing all features typical for topologically protected states. The insets in density distributions at different moments of time in Fig.~\ref{fig3}(d) show spectral density in Fourier domain, where hexagons denote Brillouin zones. Notably, such edge state populates only $\textbf{K}$ valleys during evolution. Because this state is linear it gradually expands along the edge of the insulator.

\section{Bistability of topological edge states and solitons under resonant pump}

Half-light half-matter nature of polaritons allow their resonant excitation with structured pump beams, which can be highly selective in topological microresonator arrays. To illustrate that topological edge states can be excited in system with unpaired Dirac cones, we now consider microcavity illuminated by pump with functional shape ${\mathcal{H}_\pm = h_\pm \exp(ik_x x - i\epsilon t)}$, where $\epsilon$ represents pump detuning, $k_x$ is the controllable pump momentum, while $h_\pm$ is the pump amplitude. We now take into account polariton losses described by the parameter ${\gamma=0.008}$ and polariton-polariton interactions. Searching for solutions of Eq.~(\ref{eq1}) of the form ${\psi_{\pm}=u_{\pm} \exp(ik_x x -i\epsilon t)}$, one arrives at the equation
\begin{align}\label{eq2}
	\epsilon u_{ \pm}= & - \frac{1}{2} \left[ \left( \frac{\partial}{\partial x} + i k_x \right)^2 + \frac{\partial^2}{\partial y^2}  \right] u_{ \pm} \notag \\
	& + \beta \left[ \left( \frac{\partial}{\partial x} + i k_x \right) \mp i  \frac{\partial}{\partial y}  \right]^2 u_{\mp}  \notag \\
	& + [\mathcal{R}(x, y) \pm \Omega - i \gamma]  u_{ \pm} \notag \\
	& + (|u_{ \pm}|^2+\sigma|u_{\mp}|^2) u_{ \pm}
	+ h_\pm,
\end{align}
describing nonlinear \textit{dissipative} $x$-periodic states bifurcating from representative linear modes of this system (see Fig.~\ref{fig2}) when pump is in resonance with them. We first consider ribbon configuration with zigzag-zigzag edges, whose linear spectrum is presented in Fig.~\ref{fig2}(e). Equation~(\ref{eq2}) was solved by Newton iterations method combined with plane-wave expansion approach.

\begin{figure}[h!]
	\centering
	\includegraphics[width=\columnwidth]{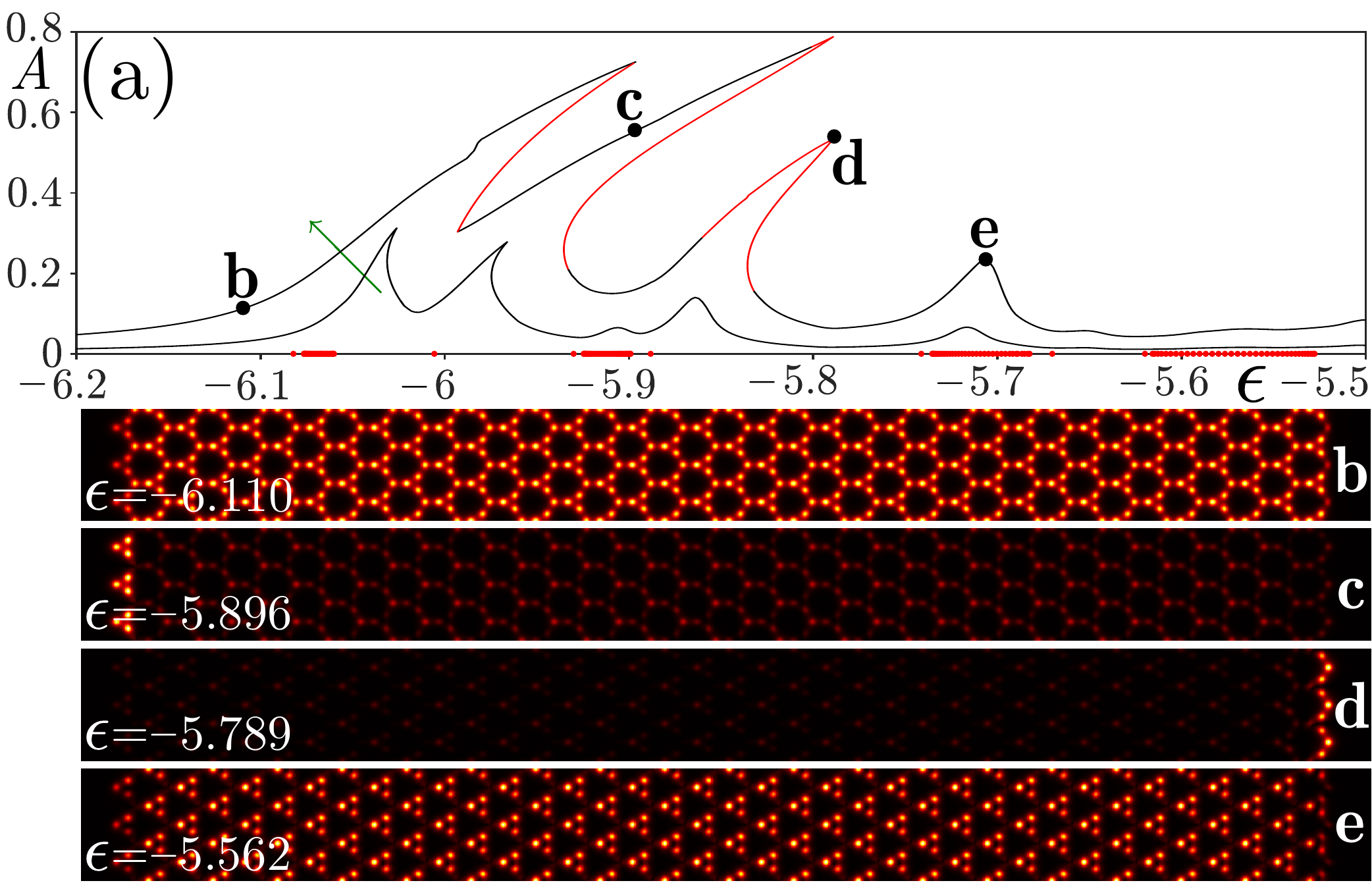}
	\caption{(a) Peak amplitude $A=\textrm{max}|\psi_-|$ of the dominating $\psi_-$ component of the excited state in ribbon configuration with zigzag-zigzag edges versus pump frequency detuning at ${k_x=0.5{\rm K}_x}$ for pump amplitudes ${h_\pm=0.001}$ and $0.004$ (increasing as indicated by the green arrow). Red and black curves represent to unstable and stable states, respectively. Red dots on the horizontal axis correspond to energies of linear modes of the array. (b-e) Density distributions of the bulk (b),(e) and edge (c),(d) states (only dominating $|\psi_-|^2$ component is shown) corresponding to the dots in resonance curves from panel (a). Parameter values are $\beta=0.43$, $\Omega=0.9$, $\gamma=0.008$.}
	\label{fig4}
\end{figure}

A major advantage of the resonant pump consists in highly selective excitation of states with different symmetry and with desired Bloch momentum $k_x$ that occurs when pump frequency detuning $\epsilon$ approaches the energy of corresponding linear mode of the system (see Fig.~\ref{fig2}). Thus, by changing both $k_x$ and $\epsilon$ one can completely change the location, structure and localization of the emerging polariton states. Coincidence of the pump detuning $\epsilon$ with eigen-energy of desired state results in resonant increase of its amplitude, provided that pump has nonzero projection on this state. We further consider linearly polarized pump with ${h_+=h_-}$. This type of excitation may be particularly fruitful in system with unpaired Dirac cones, where the domain of existence of the edge states in $k_x$ is rather limited, as shown in Fig.~\ref{fig2}(e).

The dependence of amplitude ${A=\textrm{max}|\psi_-|}$ of the excited state on pump frequency detuning $\epsilon$ is presented in Fig.~\ref{fig4}(a) for pump momentum ${k_x=0.5{\rm K}_x}$. One can observe the presence of several resonances that are tilted due to repulsive polariton-polariton interactions, that results in bistability (coexistence of three or more different types of solutions for a given $\epsilon$, with more than one solution being stable) at sufficiently large pump amplitudes. While in two outermost resonances extended bulk states are excited [Figs.~\ref{fig4}(b) and \ref{fig4}(e)], two middle resonances occur with edge states residing at the opposite edges [Figs.~\ref{fig4}(c) and \ref{fig4}(d)]. Such edge states feature small background, whose relative amplitude gradually decreases toward the tip of resonance, where most efficient excitation of the edge state occurs. Notice that stability properties for edge states are different --- the states on the right side of the ribbon are dynamically unstable on considerable part of the upper branch of resonance [see the branch labeled with point \textbf{d} in Fig.~\ref{fig4}(a)], while for state at the left edge the upper branch of resonance is stable [see the branch containing point \textbf{c} in Fig.~\ref{fig4}(a)]. Stability was tested in the frames of Eq.~(\ref{eq1}) by modeling evolution of the edge states in the presence of broadband initial perturbations. Thus, stability properties of the nonlinear edge states in this dissipative system are rather complex and depend on the structure of the edge. One can conclude that stability of the nonlinear edge state benefits from the fissioned sites, since namely stable edge states from the upper branch with state \textbf{c} are located on the boundary with fissioned sites, while the unstable states from the upper branch containing state \textbf{d} are located on the traditional zigzag boundary.

\begin{figure}[h!]
	\centering
	\includegraphics[width=\columnwidth]{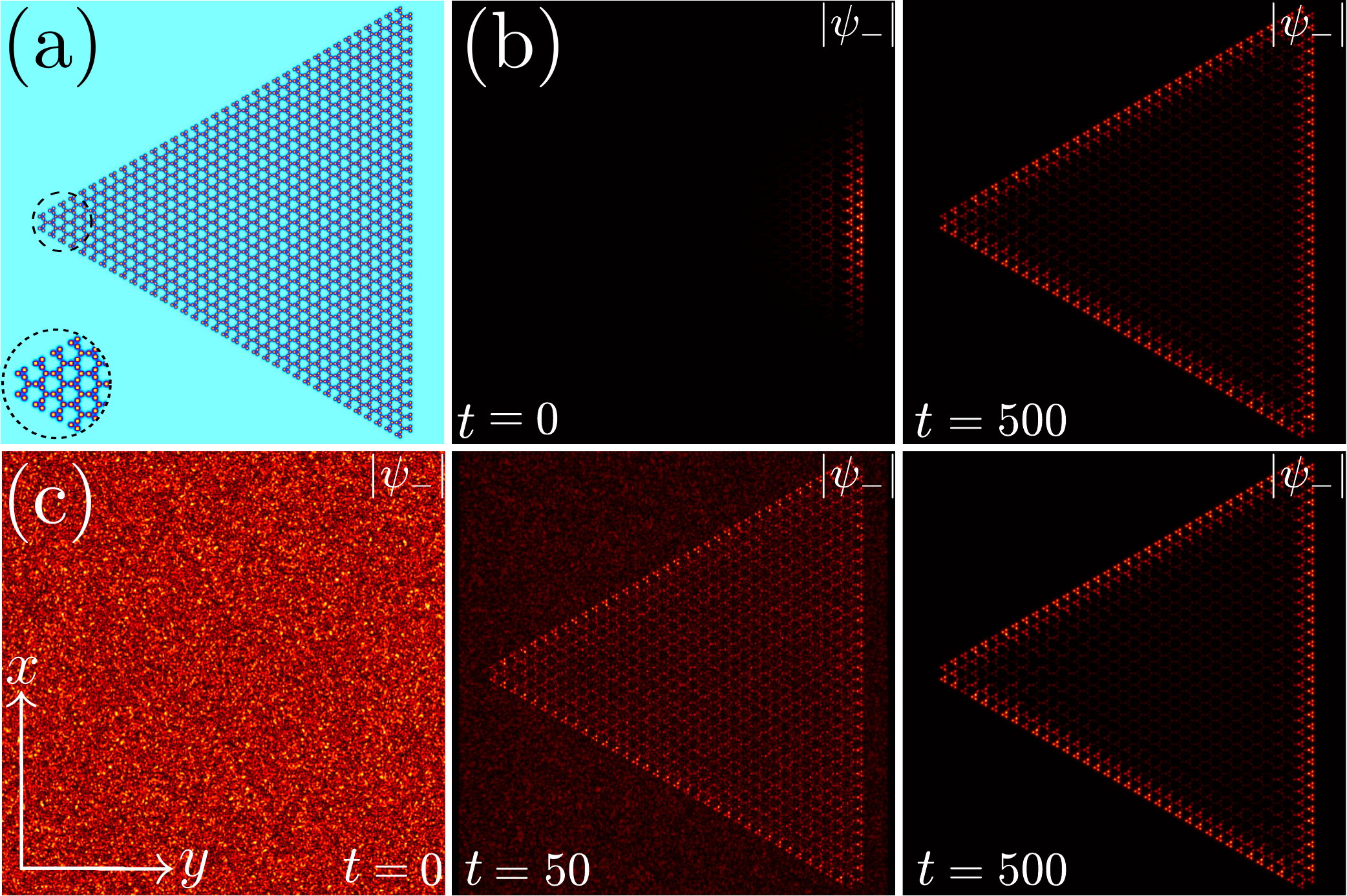}
	\caption{(a) Fissioned honeycomb array of microcavity pillars in triangular configuration with bearded edges. Inset shows the magnified sector of the structure.
		(b) Density distributions at ${t=0}$ and ${t=500}$ showing evolution of nonlinear edge state with initial Gaussian envelope under the action of uniform resonant pump. (c) Excitation of the edge state from a random noise illustrated by density distributions at ${t=0}, {t=50}$ and ${t=500}$. All distributions are shown within the ${-72\le x,y \le 72}$ window. The parameters are ${\beta=0.43}$, ${\epsilon=-5.99}$, ${\gamma=0.008}$, and ${\Omega=0.9}$. Pump amplitude $h_\pm=0.004$.}
	\label{fig5}
\end{figure}

To illustrate that edge states can be resonantly excited also in finite arrays in the regime where unpaired Dirac cones are presented in spectrum and to further test their robustness, we designed a triangular structure with bearded edge depicted in Fig.~\ref{fig5}(a) [the structure of the edge is similar to structure of the right edge of the ribbon from Fig.~\ref{fig2}(c)]. We initiate the evolution in this structure with exact nonlinear edge state at momentum ${k_x=0}$ produced by resonance with states from green edge state branch in Fig.~\ref{fig2}(f), see ${t=0}$ density distribution in Fig.~\ref{fig5}(b). The initial broad Gaussian envelope was imposed on this nonlinear edge state, so that it fits to one side of the triangular configuration, and resonant uniform pump ${\mathcal{H}_\pm = h_\pm \exp( - i\epsilon t)}$ was used. We find that upon subsequent evolution for pump detuning ${\epsilon=-5.99}$ close to the energy of linear edge state at ${k_x=0}$, the condensate uniformly populates the entire periphery of the structure, being confined near its edge [see density at ${t=500}$ in Fig.~\ref{fig5}(b)]. Moreover, similar edge state can be excited even from low-density noise [see density distribution at ${t=0}$ in Fig.~\ref{fig5}(c)], which in the presence of resonant uniform pump also evolves into stable edge state shown at ${t=500}$ in Fig.~\ref{fig5}(c) that is identical to final state from Fig.~\ref{fig5}(b). These results indicate on exceptional robustness of corresponding edge states and suggest that this system can serve as a promising platform for topological lasing applications~\cite{klembt.nature.562.552.2018, kartashov.prl.122.083902.2019, wu.sa.9.eadg4322.2023, jin.nc.16.6002.2025}.

\begin{figure}[h!]
	\centering
	\includegraphics[width=\columnwidth]{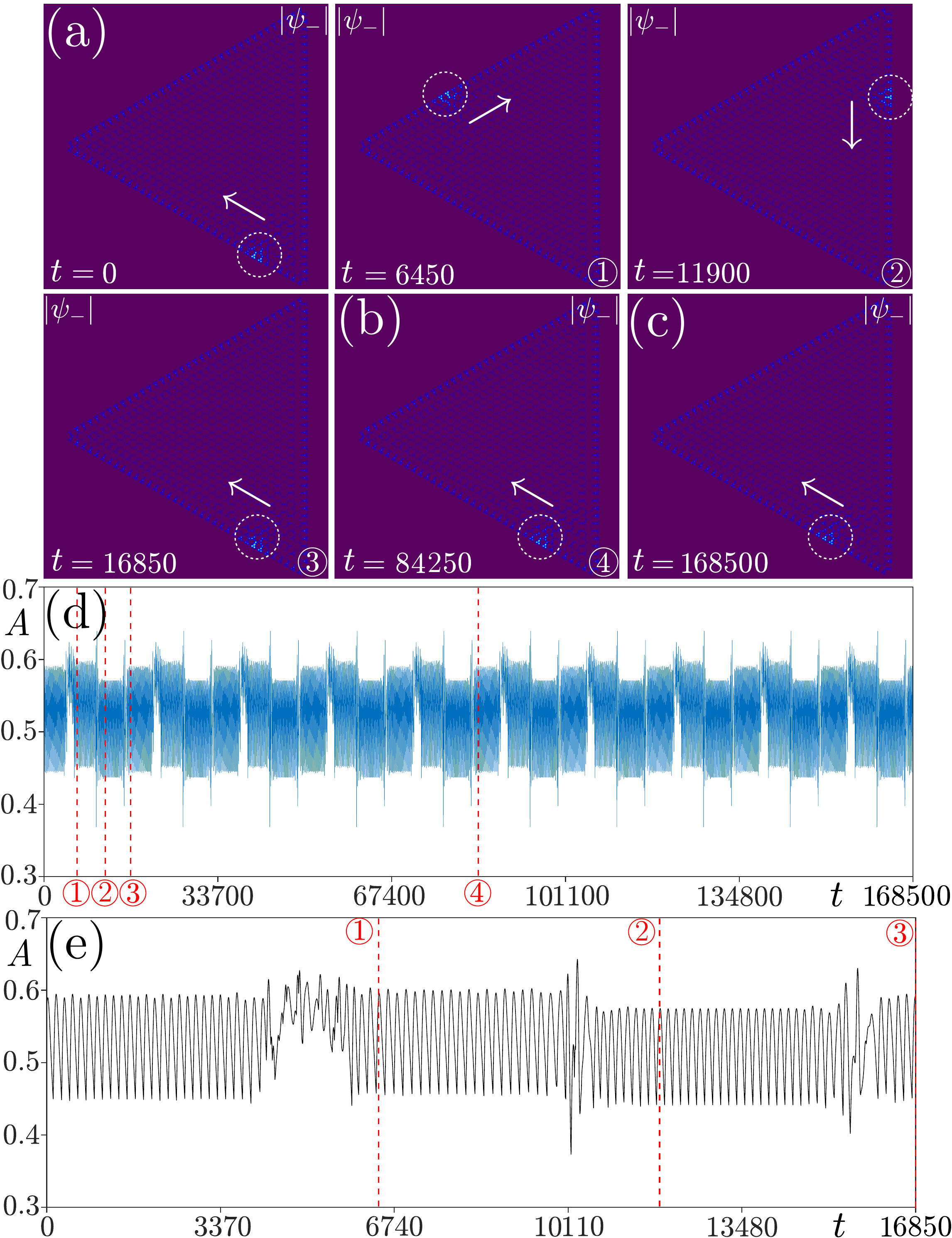}
	\caption{(a) Stable circulation of topological edge soliton along the edge of triangular array of microcavity pillars. Density distributions are shown in different moments of time. Soliton is highlighted with dashed white circle, while its circulation direction is shown by the arrow. Pump detuning ${\epsilon=-5.97}$, while other parameters are the same as in Fig.~\ref{fig5}. (d) Peak amplitude ${A=\max |\psi_-|}$ of the soliton during evolution. The vertical red dashed lines indicate the soliton at specific times that are shown in (a)-(c). (e) Same as (d) but in ${0\le t \le 16850}$.}
	\label{fig6}
\end{figure}

A remarkable result is that the finite configuration considered above can also support a new type of unidirectional dissipative soliton that can circulate along the periphery of the triangle as long as the resonant pump is on. This state resides on the nonzero (on the periphery of the triangle) background state, and it has clearly localized features, as shown in Fig.~\ref{fig6}. It can be excited by locally increasing polariton density at the edge of the structure. The soliton highlighted by white dotted circles in Fig.~\ref{fig6} moves in a clockwise direction along the edge, with the moving direction indicated by white arrows. The circulation period in this particular case with ${\epsilon=-5.97}$ is equal to ${T\sim 16850}$. The soliton passes corners of the structure, and its most important feature is that in this dissipative system, this state is never destroyed [see distributions in Figs.~\ref{fig6}(b) and \ref{fig6}(c) in time moments ${t=5T}$ and ${t=10T}$, respectively], in contrast to conservative topological edge solitons that slowly lose energy to radiation upon their motion along the edge of the insulator. To illustrate the evolution of the soliton, we display the peak amplitude ${A=\max |\psi_-|}$ versus time $t$ in Fig.~\ref{fig6}(d), which demonstrates persistent oscillations without any signatures of damping. Figure~\ref{fig6}(e) shows zoom of the above dependence in the interval from ${t=0}$ to ${t=16850}$ (one circulation period $T$) to exhibit more details of the soliton during evolution. The aperiodic disruptions in this dependence correspond to the moments of time, where soliton is circumventing the corners. For fixed pump frequency, the peak amplitude of such soliton increases with increase of pump amplitude (thus, even though the state shown in Fig.~\ref{fig6} was found accidentally, by evolving different localized inputs in the system, the whole family of stable solutions can be constructed by continuously varying pump amplitude or frequency. To the best of our knowledge, this is the first example of a unidirectional edge soliton found in a dissipative system with \textit{a resonant} pump. The existence of such solitons is tightly connected with the possibility of bistability (coexistence of nonlinear states with different amplitudes for a given detuning $\epsilon$) in this system. Given that solitons reside at the edge with fissioned sites, such fissioned sites can be essential for the formation of such solitons.

\section{Conclusion}

Summarizing, we have proposed a polariton system where unpaired Dirac cones can emerge in the linear spectrum under the breakup of both both inversion symmetry of the system and the time-reversal symmetry. The system is characterized by the coexistence of the unpaired Dirac cones and edge states in incomplete spectral gaps, which nevertheless show all features typical for topological states. The advantage of the polariton microcavity system is that the corresponding edge states can be excited resonantly and selectively by pump beams with specially adjusted frequency and spatial structure in both ribbon and finite triangular configurations. Despite the absence of a complete spectral gap, edge states show clear unidirectional signatures, without backscattering or inter-valley scattering in corners or on defects of the structure. This suggests that topological properties can be encountered in a broad class of systems with unpaired Dirac cones and that they can lead to unconventional dynamical regimes with coexisting qualitatively different types of evolutions of edge excitations. Fissioned honeycomb arrays of microcavity pillars proposed here also suggest a promising platform for the exploration of higher-order topological phases, effects associated with fractality or global topological disclinations and defects introduced into the structure, and flat-band physics in dissipative polariton systems~\cite{xie.nrp.3.520.2021, lin.nrp.5.483.2023, ren.light.12.194.2023, zhong.light.13.264.2024, kompanets.am.37.2500556.2025}.

\begin{acknowledgments}		
This work was supported by the Natural Science Basic Research Program of Shaanxi Province (2024JC-JCQN-06, 2025JC-QYCX-006), the National Natural Science Foundation of China (12474337), the Key Research and Development Program of Shaanxi Province (2025CY-YBXM-037), the Postdoctoral Research Project of Shaanxi Province (2023BSHYDZZ14), the Sichuan Science and Technology Program (2025ZNSFSC1458), the Russian Science Foundation (grant 24-12-00167) and partially by the project FFUU-2024-0003 of the Institute of Spectroscopy of Russian Academy of Sciences.
\end{acknowledgments}

%\bibliographystyle{myprl}
%\bibliography{my_library,my_library4}

\begin{thebibliography}{70}%
	\makeatletter
	\providecommand \@ifxundefined [1]{%
		\@ifx{#1\undefined}
	}%
	\providecommand \@ifnum [1]{%
		\ifnum #1\expandafter \@firstoftwo
		\else \expandafter \@secondoftwo
		\fi
	}%
	\providecommand \@ifx [1]{%
		\ifx #1\expandafter \@firstoftwo
		\else \expandafter \@secondoftwo
		\fi
	}%
	\providecommand \natexlab [1]{#1}%
	\providecommand \enquote  [1]{``#1''}%
	\providecommand \bibnamefont  [1]{#1}%
	\providecommand \bibfnamefont [1]{#1}%
	\providecommand \citenamefont [1]{#1}%
	\providecommand \href@noop [0]{\@secondoftwo}%
	\providecommand \href [0]{\begingroup \@sanitize@url \@href}%
	\providecommand \@href[1]{\@@startlink{#1}\@@href}%
	\providecommand \@@href[1]{\endgroup#1\@@endlink}%
	\providecommand \@sanitize@url [0]{\catcode `\\12\catcode `\$12\catcode
		`\&12\catcode `\#12\catcode `\^12\catcode `\_12\catcode `\%12\relax}%
	\providecommand \@@startlink[1]{}%
	\providecommand \@@endlink[0]{}%
	\providecommand \url  [0]{\begingroup\@sanitize@url \@url }%
	\providecommand \@url [1]{\endgroup\@href {#1}{\urlprefix }}%
	\providecommand \urlprefix  [0]{URL }%
	\providecommand \Eprint [0]{\href }%
	\providecommand \doibase [0]{https://doi.org/}%
	\providecommand \selectlanguage [0]{\@gobble}%
	\providecommand \bibinfo  [0]{\@secondoftwo}%
	\providecommand \bibfield  [0]{\@secondoftwo}%
	\providecommand \translation [1]{[#1]}%
	\providecommand \BibitemOpen [0]{}%
	\providecommand \bibitemStop [0]{}%
	\providecommand \bibitemNoStop [0]{.\EOS\space}%
	\providecommand \EOS [0]{\spacefactor3000\relax}%
	\providecommand \BibitemShut  [1]{\csname bibitem#1\endcsname}%
	\let\auto@bib@innerbib\@empty
	%</preamble>
	\bibitem [{\citenamefont {Haldane}(1988)}]{haldane.prl.61.2015.1988}%
	\BibitemOpen
	\bibfield  {author} {\bibinfo {author} {\bibfnamefont {F.~D.~M.}\
			\bibnamefont {Haldane}},\ }\bibfield  {title} {\bibinfo {title} {Model for a
			quantum {Hall} effect without {Landau} levels: Condensed-matter realization
			of the ``parity anomaly''},\ }\href
	{https://doi.org/10.1103/PhysRevLett.61.2015} {\bibfield  {journal} {\bibinfo
			{journal} {Phys. Rev. Lett.}\ }\textbf {\bibinfo {volume} {61}},\ \bibinfo
		{pages} {2015} (\bibinfo {year} {1988})}\BibitemShut {NoStop}%
	\bibitem [{\citenamefont {Jotzu}\ \emph {et~al.}(2014)\citenamefont {Jotzu},
		\citenamefont {Messer}, \citenamefont {Desbuquois}, \citenamefont {Lebrat},
		\citenamefont {Uehlinger}, \citenamefont {Greif},\ and\ \citenamefont
		{Esslinger}}]{jotzu.nature.515.237.2014}%
	\BibitemOpen
	\bibfield  {author} {\bibinfo {author} {\bibfnamefont {G.}~\bibnamefont
			{Jotzu}}, \bibinfo {author} {\bibfnamefont {M.}~\bibnamefont {Messer}},
		\bibinfo {author} {\bibfnamefont {R.}~\bibnamefont {Desbuquois}}, \bibinfo
		{author} {\bibfnamefont {M.}~\bibnamefont {Lebrat}}, \bibinfo {author}
		{\bibfnamefont {T.}~\bibnamefont {Uehlinger}}, \bibinfo {author}
		{\bibfnamefont {D.}~\bibnamefont {Greif}},\ and\ \bibinfo {author}
		{\bibfnamefont {T.}~\bibnamefont {Esslinger}},\ }\bibfield  {title} {\bibinfo
		{title} {Experimental realisation of the topological {H}aldane model},\
	}\href {https://doi.org/10.1038/nature13915} {\bibfield  {journal} {\bibinfo
			{journal} {Nature}\ }\textbf {\bibinfo {volume} {515}},\ \bibinfo {pages}
		{237} (\bibinfo {year} {2014})}\BibitemShut {NoStop}%
	\bibitem [{\citenamefont {Lu}\ \emph {et~al.}(2014)\citenamefont {Lu},
		\citenamefont {Joannopoulos},\ and\ \citenamefont
		{Solja{\v{c}}i{\'c}}}]{lu.np.8.821.2014}%
	\BibitemOpen
	\bibfield  {author} {\bibinfo {author} {\bibfnamefont {L.}~\bibnamefont
			{Lu}}, \bibinfo {author} {\bibfnamefont {J.~D.}\ \bibnamefont
			{Joannopoulos}},\ and\ \bibinfo {author} {\bibfnamefont {M.}~\bibnamefont
			{Solja{\v{c}}i{\'c}}},\ }\bibfield  {title} {\bibinfo {title} {Topological
			photonics},\ }\href {https://doi.org/10.1038/nphoton.2014.248} {\bibfield
		{journal} {\bibinfo  {journal} {Nat. Photon.}\ }\textbf {\bibinfo {volume}
			{8}},\ \bibinfo {pages} {821} (\bibinfo {year} {2014})}\BibitemShut {NoStop}%
	\bibitem [{\citenamefont {Ozawa}\ \emph {et~al.}(2019)\citenamefont {Ozawa},
		\citenamefont {Price}, \citenamefont {Amo}, \citenamefont {Goldman},
		\citenamefont {Hafezi}, \citenamefont {Lu}, \citenamefont {Rechtsman},
		\citenamefont {Schuster}, \citenamefont {Simon}, \citenamefont {Zilberberg},\
		and\ \citenamefont {Carusotto}}]{ozawa.rmp.91.015006.2019}%
	\BibitemOpen
	\bibfield  {author} {\bibinfo {author} {\bibfnamefont {T.}~\bibnamefont
			{Ozawa}}, \bibinfo {author} {\bibfnamefont {H.~M.}\ \bibnamefont {Price}},
		\bibinfo {author} {\bibfnamefont {A.}~\bibnamefont {Amo}}, \bibinfo {author}
		{\bibfnamefont {N.}~\bibnamefont {Goldman}}, \bibinfo {author} {\bibfnamefont
			{M.}~\bibnamefont {Hafezi}}, \bibinfo {author} {\bibfnamefont
			{L.}~\bibnamefont {Lu}}, \bibinfo {author} {\bibfnamefont {M.~C.}\
			\bibnamefont {Rechtsman}}, \bibinfo {author} {\bibfnamefont {D.}~\bibnamefont
			{Schuster}}, \bibinfo {author} {\bibfnamefont {J.}~\bibnamefont {Simon}},
		\bibinfo {author} {\bibfnamefont {O.}~\bibnamefont {Zilberberg}},\ and\
		\bibinfo {author} {\bibfnamefont {I.}~\bibnamefont {Carusotto}},\ }\bibfield
	{title} {\bibinfo {title} {Topological photonics},\ }\href
	{https://doi.org/10.1103/RevModPhys.91.015006} {\bibfield  {journal}
		{\bibinfo  {journal} {Rev. Mod. Phys.}\ }\textbf {\bibinfo {volume} {91}},\
		\bibinfo {pages} {015006} (\bibinfo {year} {2019})}\BibitemShut {NoStop}%
	\bibitem [{\citenamefont {Zhang}\ \emph {et~al.}(2023)\citenamefont {Zhang},
		\citenamefont {Zangeneh-Nejad}, \citenamefont {Chen}, \citenamefont {Lu},\
		and\ \citenamefont {Christensen}}]{zhang.nature.618.687.2023}%
	\BibitemOpen
	\bibfield  {author} {\bibinfo {author} {\bibfnamefont {X.}~\bibnamefont
			{Zhang}}, \bibinfo {author} {\bibfnamefont {F.}~\bibnamefont
			{Zangeneh-Nejad}}, \bibinfo {author} {\bibfnamefont {Z.-G.}\ \bibnamefont
			{Chen}}, \bibinfo {author} {\bibfnamefont {M.-H.}\ \bibnamefont {Lu}},\ and\
		\bibinfo {author} {\bibfnamefont {J.}~\bibnamefont {Christensen}},\
	}\bibfield  {title} {\bibinfo {title} {A second wave of topological phenomena
			in photonics and acoustics},\ }\href
	{https://doi.org/10.1038/s41586-023-06163-9} {\bibfield  {journal} {\bibinfo
			{journal} {Nature}\ }\textbf {\bibinfo {volume} {618}},\ \bibinfo {pages}
		{687} (\bibinfo {year} {2023})}\BibitemShut {NoStop}%
	\bibitem [{\citenamefont {Leykam}\ \emph {et~al.}(2026)\citenamefont {Leykam},
		\citenamefont {Xue}, \citenamefont {Zhang},\ and\ \citenamefont
		{Chong}}]{leykam.nrp.8.55.2026}%
	\BibitemOpen
	\bibfield  {author} {\bibinfo {author} {\bibfnamefont {D.}~\bibnamefont
			{Leykam}}, \bibinfo {author} {\bibfnamefont {H.}~\bibnamefont {Xue}},
		\bibinfo {author} {\bibfnamefont {B.}~\bibnamefont {Zhang}},\ and\ \bibinfo
		{author} {\bibfnamefont {Y.~D.}\ \bibnamefont {Chong}},\ }\bibfield  {title}
	{\bibinfo {title} {Limitations and possibilities of topological photonics},\
	}\href {https://doi.org/10.1038/s42254-025-00889-3} {\bibfield  {journal}
		{\bibinfo  {journal} {Nat. Rev. Phys.}\ }\textbf {\bibinfo {volume} {8}},\
		\bibinfo {pages} {55} (\bibinfo {year} {2026})}\BibitemShut {NoStop}%
	\bibitem [{\citenamefont {Rechtsman}\ \emph {et~al.}(2013)\citenamefont
		{Rechtsman}, \citenamefont {Zeuner}, \citenamefont {Plotnik}, \citenamefont
		{Lumer}, \citenamefont {Podolsky}, \citenamefont {Dreisow}, \citenamefont
		{Nolte}, \citenamefont {Segev},\ and\ \citenamefont
		{Szameit}}]{rechtsman.nature.496.196.2013}%
	\BibitemOpen
	\bibfield  {author} {\bibinfo {author} {\bibfnamefont {M.~C.}\ \bibnamefont
			{Rechtsman}}, \bibinfo {author} {\bibfnamefont {J.~M.}\ \bibnamefont
			{Zeuner}}, \bibinfo {author} {\bibfnamefont {Y.}~\bibnamefont {Plotnik}},
		\bibinfo {author} {\bibfnamefont {Y.}~\bibnamefont {Lumer}}, \bibinfo
		{author} {\bibfnamefont {D.}~\bibnamefont {Podolsky}}, \bibinfo {author}
		{\bibfnamefont {F.}~\bibnamefont {Dreisow}}, \bibinfo {author} {\bibfnamefont
			{S.}~\bibnamefont {Nolte}}, \bibinfo {author} {\bibfnamefont
			{M.}~\bibnamefont {Segev}},\ and\ \bibinfo {author} {\bibfnamefont
			{A.}~\bibnamefont {Szameit}},\ }\bibfield  {title} {\bibinfo {title}
		{Photonic {F}loquet topological insulators},\ }\href
	{https://doi.org/10.1038/nature12066} {\bibfield  {journal} {\bibinfo
			{journal} {Nature}\ }\textbf {\bibinfo {volume} {496}},\ \bibinfo {pages}
		{196} (\bibinfo {year} {2013})}\BibitemShut {NoStop}%
	\bibitem [{\citenamefont {Leykam}\ \emph {et~al.}(2016)\citenamefont {Leykam},
		\citenamefont {Rechtsman},\ and\ \citenamefont
		{Chong}}]{leykam.prl.117.013902.2016}%
	\BibitemOpen
	\bibfield  {author} {\bibinfo {author} {\bibfnamefont {D.}~\bibnamefont
			{Leykam}}, \bibinfo {author} {\bibfnamefont {M.~C.}\ \bibnamefont
			{Rechtsman}},\ and\ \bibinfo {author} {\bibfnamefont {Y.~D.}\ \bibnamefont
			{Chong}},\ }\bibfield  {title} {\bibinfo {title} {Anomalous topological
			phases and unpaired {Dirac} cones in photonic {F}loquet topological
			insulators},\ }\href {https://doi.org/10.1103/PhysRevLett.117.013902}
	{\bibfield  {journal} {\bibinfo  {journal} {Phys. Rev. Lett.}\ }\textbf
		{\bibinfo {volume} {117}},\ \bibinfo {pages} {013902} (\bibinfo {year}
		{2016})}\BibitemShut {NoStop}%
	\bibitem [{\citenamefont {Zhong}\ \emph
		{et~al.}(2024{\natexlab{a}})\citenamefont {Zhong}, \citenamefont {Kartashov},
		\citenamefont {Li}, \citenamefont {Li},\ and\ \citenamefont
		{Zhang}}]{zhong.pr.12.2078.2024}%
	\BibitemOpen
	\bibfield  {author} {\bibinfo {author} {\bibfnamefont {H.}~\bibnamefont
			{Zhong}}, \bibinfo {author} {\bibfnamefont {Y.~V.}\ \bibnamefont
			{Kartashov}}, \bibinfo {author} {\bibfnamefont {Y.}~\bibnamefont {Li}},
		\bibinfo {author} {\bibfnamefont {M.}~\bibnamefont {Li}},\ and\ \bibinfo
		{author} {\bibfnamefont {Y.}~\bibnamefont {Zhang}},\ }\bibfield  {title}
	{\bibinfo {title} {Topological edge states in a photonic {Floquet} insulator
			with unpaired {Dirac} cones},\ }\href {https://doi.org/10.1364/PRJ.524824}
	{\bibfield  {journal} {\bibinfo  {journal} {Photon. Res.}\ }\textbf {\bibinfo
			{volume} {12}},\ \bibinfo {pages} {2078} (\bibinfo {year}
		{2024}{\natexlab{a}})}\BibitemShut {NoStop}%
	\bibitem [{\citenamefont {Xiao}\ \emph {et~al.}(2007)\citenamefont {Xiao},
		\citenamefont {Yao},\ and\ \citenamefont {Niu}}]{xiao.prl.99.236809.2007}%
	\BibitemOpen
	\bibfield  {author} {\bibinfo {author} {\bibfnamefont {D.}~\bibnamefont
			{Xiao}}, \bibinfo {author} {\bibfnamefont {W.}~\bibnamefont {Yao}},\ and\
		\bibinfo {author} {\bibfnamefont {Q.}~\bibnamefont {Niu}},\ }\bibfield
	{title} {\bibinfo {title} {Valley-contrasting physics in graphene: Magnetic
			moment and topological transport},\ }\href
	{https://doi.org/10.1103/PhysRevLett.99.236809} {\bibfield  {journal}
		{\bibinfo  {journal} {Phys. Rev. Lett.}\ }\textbf {\bibinfo {volume} {99}},\
		\bibinfo {pages} {236809} (\bibinfo {year} {2007})}\BibitemShut {NoStop}%
	\bibitem [{\citenamefont {Noh}\ \emph {et~al.}(2018)\citenamefont {Noh},
		\citenamefont {Huang}, \citenamefont {Chen},\ and\ \citenamefont
		{Rechtsman}}]{noh.prl.120.063902.2018}%
	\BibitemOpen
	\bibfield  {author} {\bibinfo {author} {\bibfnamefont {J.}~\bibnamefont
			{Noh}}, \bibinfo {author} {\bibfnamefont {S.}~\bibnamefont {Huang}}, \bibinfo
		{author} {\bibfnamefont {K.~P.}\ \bibnamefont {Chen}},\ and\ \bibinfo
		{author} {\bibfnamefont {M.~C.}\ \bibnamefont {Rechtsman}},\ }\bibfield
	{title} {\bibinfo {title} {Observation of photonic topological valley {H}all
			edge states},\ }\href {https://doi.org/10.1103/PhysRevLett.120.063902}
	{\bibfield  {journal} {\bibinfo  {journal} {Phys. Rev. Lett.}\ }\textbf
		{\bibinfo {volume} {120}},\ \bibinfo {pages} {063902} (\bibinfo {year}
		{2018})}\BibitemShut {NoStop}%
	\bibitem [{\citenamefont {Xue}\ \emph {et~al.}(2021)\citenamefont {Xue},
		\citenamefont {Yang},\ and\ \citenamefont {Zhang}}]{xue.apr.2.2100013.2021}%
	\BibitemOpen
	\bibfield  {author} {\bibinfo {author} {\bibfnamefont {H.}~\bibnamefont
			{Xue}}, \bibinfo {author} {\bibfnamefont {Y.}~\bibnamefont {Yang}},\ and\
		\bibinfo {author} {\bibfnamefont {B.}~\bibnamefont {Zhang}},\ }\bibfield
	{title} {\bibinfo {title} {Topological valley photonics: Physics and device
			applications},\ }\href
	{https://doi.org/https://doi.org/10.1002/adpr.202100013} {\bibfield
		{journal} {\bibinfo  {journal} {Adv. Photon. Research}\ }\textbf {\bibinfo
			{volume} {2}},\ \bibinfo {pages} {2100013} (\bibinfo {year}
		{2021})}\BibitemShut {NoStop}%
	\bibitem [{\citenamefont {Tang}\ \emph
		{et~al.}(2022{\natexlab{a}})\citenamefont {Tang}, \citenamefont {He},
		\citenamefont {Shi}, \citenamefont {Liu}, \citenamefont {Chen},\ and\
		\citenamefont {Dong}}]{tang.lpr.1.2100300.2022}%
	\BibitemOpen
	\bibfield  {author} {\bibinfo {author} {\bibfnamefont {G.-J.}\ \bibnamefont
			{Tang}}, \bibinfo {author} {\bibfnamefont {X.-T.}\ \bibnamefont {He}},
		\bibinfo {author} {\bibfnamefont {F.-L.}\ \bibnamefont {Shi}}, \bibinfo
		{author} {\bibfnamefont {J.-W.}\ \bibnamefont {Liu}}, \bibinfo {author}
		{\bibfnamefont {X.-D.}\ \bibnamefont {Chen}},\ and\ \bibinfo {author}
		{\bibfnamefont {J.-W.}\ \bibnamefont {Dong}},\ }\bibfield  {title} {\bibinfo
		{title} {Topological photonic crystals: Physics, designs, and applications},\
	}\href {https://doi.org/10.1002/lpor.202100300} {\bibfield  {journal}
		{\bibinfo  {journal} {Laser Photon. Rev.}\ }\textbf {\bibinfo {volume}
			{16}},\ \bibinfo {pages} {2100300} (\bibinfo {year}
		{2022}{\natexlab{a}})}\BibitemShut {NoStop}%
	\bibitem [{\citenamefont {Zhou}\ \emph {et~al.}(2018)\citenamefont {Zhou},
		\citenamefont {Leykam}, \citenamefont {Chattopadhyay}, \citenamefont
		{Khanikaev},\ and\ \citenamefont {Chong}}]{zhou.prb.98.205115.2018}%
	\BibitemOpen
	\bibfield  {author} {\bibinfo {author} {\bibfnamefont {X.}~\bibnamefont
			{Zhou}}, \bibinfo {author} {\bibfnamefont {D.}~\bibnamefont {Leykam}},
		\bibinfo {author} {\bibfnamefont {U.}~\bibnamefont {Chattopadhyay}}, \bibinfo
		{author} {\bibfnamefont {A.~B.}\ \bibnamefont {Khanikaev}},\ and\ \bibinfo
		{author} {\bibfnamefont {Y.~D.}\ \bibnamefont {Chong}},\ }\bibfield  {title}
	{\bibinfo {title} {Realization of a magneto-optical near-zero index medium by
			an unpaired {Dirac} point},\ }\href
	{https://doi.org/10.1103/PhysRevB.98.205115} {\bibfield  {journal} {\bibinfo
			{journal} {Phys. Rev. B}\ }\textbf {\bibinfo {volume} {98}},\ \bibinfo
		{pages} {205115} (\bibinfo {year} {2018})}\BibitemShut {NoStop}%
	\bibitem [{\citenamefont {Xue}\ \emph {et~al.}(2020)\citenamefont {Xue},
		\citenamefont {Wang}, \citenamefont {Zhang},\ and\ \citenamefont
		{Chong}}]{xue.prl.124.236403.2020}%
	\BibitemOpen
	\bibfield  {author} {\bibinfo {author} {\bibfnamefont {H.}~\bibnamefont
			{Xue}}, \bibinfo {author} {\bibfnamefont {Q.}~\bibnamefont {Wang}}, \bibinfo
		{author} {\bibfnamefont {B.}~\bibnamefont {Zhang}},\ and\ \bibinfo {author}
		{\bibfnamefont {Y.~D.}\ \bibnamefont {Chong}},\ }\bibfield  {title} {\bibinfo
		{title} {Non-{Hermitian Dirac} cones},\ }\href
	{https://doi.org/10.1103/PhysRevLett.124.236403} {\bibfield  {journal}
		{\bibinfo  {journal} {Phys. Rev. Lett.}\ }\textbf {\bibinfo {volume} {124}},\
		\bibinfo {pages} {236403} (\bibinfo {year} {2020})}\BibitemShut {NoStop}%
	\bibitem [{\citenamefont {Wang}\ \emph {et~al.}(2023)\citenamefont {Wang},
		\citenamefont {Wang}, \citenamefont {Liang}, \citenamefont {Zhu},
		\citenamefont {Fan}, \citenamefont {Lin}, \citenamefont {Li}, \citenamefont
		{Zhang}, \citenamefont {Luan}, \citenamefont {Poo}, \citenamefont {Jiang},\
		and\ \citenamefont {Guo}}]{wang.nc.14.4457.2023}%
	\BibitemOpen
	\bibfield  {author} {\bibinfo {author} {\bibfnamefont {Y.}~\bibnamefont
			{Wang}}, \bibinfo {author} {\bibfnamefont {H.-X.}\ \bibnamefont {Wang}},
		\bibinfo {author} {\bibfnamefont {L.}~\bibnamefont {Liang}}, \bibinfo
		{author} {\bibfnamefont {W.}~\bibnamefont {Zhu}}, \bibinfo {author}
		{\bibfnamefont {L.}~\bibnamefont {Fan}}, \bibinfo {author} {\bibfnamefont
			{Z.-K.}\ \bibnamefont {Lin}}, \bibinfo {author} {\bibfnamefont
			{F.}~\bibnamefont {Li}}, \bibinfo {author} {\bibfnamefont {X.}~\bibnamefont
			{Zhang}}, \bibinfo {author} {\bibfnamefont {P.-G.}\ \bibnamefont {Luan}},
		\bibinfo {author} {\bibfnamefont {Y.}~\bibnamefont {Poo}}, \bibinfo {author}
		{\bibfnamefont {J.-H.}\ \bibnamefont {Jiang}},\ and\ \bibinfo {author}
		{\bibfnamefont {G.-Y.}\ \bibnamefont {Guo}},\ }\bibfield  {title} {\bibinfo
		{title} {Hybrid topological photonic crystals},\ }\href
	{https://doi.org/10.1038/s41467-023-40172-6} {\bibfield  {journal} {\bibinfo
			{journal} {Nat. Commun.}\ }\textbf {\bibinfo {volume} {14}},\ \bibinfo
		{pages} {4457} (\bibinfo {year} {2023})}\BibitemShut {NoStop}%
	\bibitem [{\citenamefont {Liu}\ \emph {et~al.}(2020{\natexlab{a}})\citenamefont
		{Liu}, \citenamefont {Zhou}, \citenamefont {Yang}, \citenamefont {Xue},
		\citenamefont {Ren}, \citenamefont {Lin}, \citenamefont {Sun}, \citenamefont
		{Bi}, \citenamefont {Chong},\ and\ \citenamefont {Zhang}}]{liu.nc.1873.2020}%
	\BibitemOpen
	\bibfield  {author} {\bibinfo {author} {\bibfnamefont {G.-G.}\ \bibnamefont
			{Liu}}, \bibinfo {author} {\bibfnamefont {P.}~\bibnamefont {Zhou}}, \bibinfo
		{author} {\bibfnamefont {Y.}~\bibnamefont {Yang}}, \bibinfo {author}
		{\bibfnamefont {H.}~\bibnamefont {Xue}}, \bibinfo {author} {\bibfnamefont
			{X.}~\bibnamefont {Ren}}, \bibinfo {author} {\bibfnamefont {X.}~\bibnamefont
			{Lin}}, \bibinfo {author} {\bibfnamefont {H.-x.}\ \bibnamefont {Sun}},
		\bibinfo {author} {\bibfnamefont {L.}~\bibnamefont {Bi}}, \bibinfo {author}
		{\bibfnamefont {Y.}~\bibnamefont {Chong}},\ and\ \bibinfo {author}
		{\bibfnamefont {B.}~\bibnamefont {Zhang}},\ }\bibfield  {title} {\bibinfo
		{title} {Observation of an unpaired photonic {Dirac} point},\ }\href
	{https://doi.org/10.1038/s41467-020-15801-z} {\bibfield  {journal} {\bibinfo
			{journal} {Nat. Commun.}\ }\textbf {\bibinfo {volume} {11}},\ \bibinfo
		{pages} {1873} (\bibinfo {year} {2020}{\natexlab{a}})}\BibitemShut {NoStop}%
	\bibitem [{\citenamefont {Klembt}\ \emph {et~al.}(2018)\citenamefont {Klembt},
		\citenamefont {Harder}, \citenamefont {Egorov}, \citenamefont {Winkler},
		\citenamefont {Ge}, \citenamefont {Bandres}, \citenamefont {Emmerling},
		\citenamefont {Worschech}, \citenamefont {Liew}, \citenamefont {Segev},
		\citenamefont {Schneider},\ and\ \citenamefont
		{H\"{o}fling}}]{klembt.nature.562.552.2018}%
	\BibitemOpen
	\bibfield  {author} {\bibinfo {author} {\bibfnamefont {S.}~\bibnamefont
			{Klembt}}, \bibinfo {author} {\bibfnamefont {T.~H.}\ \bibnamefont {Harder}},
		\bibinfo {author} {\bibfnamefont {O.~A.}\ \bibnamefont {Egorov}}, \bibinfo
		{author} {\bibfnamefont {K.}~\bibnamefont {Winkler}}, \bibinfo {author}
		{\bibfnamefont {R.}~\bibnamefont {Ge}}, \bibinfo {author} {\bibfnamefont
			{M.~A.}\ \bibnamefont {Bandres}}, \bibinfo {author} {\bibfnamefont
			{M.}~\bibnamefont {Emmerling}}, \bibinfo {author} {\bibfnamefont
			{L.}~\bibnamefont {Worschech}}, \bibinfo {author} {\bibfnamefont {T.~C.~H.}\
			\bibnamefont {Liew}}, \bibinfo {author} {\bibfnamefont {M.}~\bibnamefont
			{Segev}}, \bibinfo {author} {\bibfnamefont {C.}~\bibnamefont {Schneider}},\
		and\ \bibinfo {author} {\bibfnamefont {S.}~\bibnamefont {H\"{o}fling}},\
	}\bibfield  {title} {\bibinfo {title} {Exciton-polariton topological
			insulator},\ }\href {https://doi.org/10.1038/s41586-018-0601-5} {\bibfield
		{journal} {\bibinfo  {journal} {Nature}\ }\textbf {\bibinfo {volume} {562}},\
		\bibinfo {pages} {552} (\bibinfo {year} {2018})}\BibitemShut {NoStop}%
	\bibitem [{\citenamefont {Nalitov}\ \emph {et~al.}(2015)\citenamefont
		{Nalitov}, \citenamefont {Solnyshkov},\ and\ \citenamefont
		{Malpuech}}]{nalitov.prl.114.116401.2015}%
	\BibitemOpen
	\bibfield  {author} {\bibinfo {author} {\bibfnamefont {A.~V.}\ \bibnamefont
			{Nalitov}}, \bibinfo {author} {\bibfnamefont {D.~D.}\ \bibnamefont
			{Solnyshkov}},\ and\ \bibinfo {author} {\bibfnamefont {G.}~\bibnamefont
			{Malpuech}},\ }\bibfield  {title} {\bibinfo {title} {Polariton $\mathbb{Z}$
			topological insulator},\ }\href
	{https://doi.org/10.1103/PhysRevLett.114.116401} {\bibfield  {journal}
		{\bibinfo  {journal} {Phys. Rev. Lett.}\ }\textbf {\bibinfo {volume} {114}},\
		\bibinfo {pages} {116401} (\bibinfo {year} {2015})}\BibitemShut {NoStop}%
	\bibitem [{\citenamefont {Bardyn}\ \emph {et~al.}(2015)\citenamefont {Bardyn},
		\citenamefont {Karzig}, \citenamefont {Refael},\ and\ \citenamefont
		{Liew}}]{bardyn.prb.91.161413.2015}%
	\BibitemOpen
	\bibfield  {author} {\bibinfo {author} {\bibfnamefont {C.-E.}\ \bibnamefont
			{Bardyn}}, \bibinfo {author} {\bibfnamefont {T.}~\bibnamefont {Karzig}},
		\bibinfo {author} {\bibfnamefont {G.}~\bibnamefont {Refael}},\ and\ \bibinfo
		{author} {\bibfnamefont {T.~C.~H.}\ \bibnamefont {Liew}},\ }\bibfield
	{title} {\bibinfo {title} {Topological polaritons and excitons in
			garden-variety systems},\ }\href {https://doi.org/10.1103/PhysRevB.91.161413}
	{\bibfield  {journal} {\bibinfo  {journal} {Phys. Rev. B}\ }\textbf {\bibinfo
			{volume} {91}},\ \bibinfo {pages} {161413} (\bibinfo {year}
		{2015})}\BibitemShut {NoStop}%
	\bibitem [{\citenamefont {Karzig}\ \emph {et~al.}(2015)\citenamefont {Karzig},
		\citenamefont {Bardyn}, \citenamefont {Lindner},\ and\ \citenamefont
		{Refael}}]{karzig.prx.5.031001.2015}%
	\BibitemOpen
	\bibfield  {author} {\bibinfo {author} {\bibfnamefont {T.}~\bibnamefont
			{Karzig}}, \bibinfo {author} {\bibfnamefont {C.-E.}\ \bibnamefont {Bardyn}},
		\bibinfo {author} {\bibfnamefont {N.~H.}\ \bibnamefont {Lindner}},\ and\
		\bibinfo {author} {\bibfnamefont {G.}~\bibnamefont {Refael}},\ }\bibfield
	{title} {\bibinfo {title} {Topological polaritons},\ }\href
	{https://doi.org/10.1103/PhysRevX.5.031001} {\bibfield  {journal} {\bibinfo
			{journal} {Phys. Rev. X}\ }\textbf {\bibinfo {volume} {5}},\ \bibinfo {pages}
		{031001} (\bibinfo {year} {2015})}\BibitemShut {NoStop}%
	\bibitem [{\citenamefont {Mili{\'{c}}evi{\'{c}}}\ \emph
		{et~al.}(2015)\citenamefont {Mili{\'{c}}evi{\'{c}}}, \citenamefont {Ozawa},
		\citenamefont {Andreakou}, \citenamefont {Carusotto}, \citenamefont
		{Jacqmin}, \citenamefont {Galopin}, \citenamefont {Lema{\^{\i}}tre},
		\citenamefont {Gratiet}, \citenamefont {Sagnes}, \citenamefont {Bloch},\ and\
		\citenamefont {Amo}}]{milivic.2d.2.034012.2015}%
	\BibitemOpen
	\bibfield  {author} {\bibinfo {author} {\bibfnamefont {M.}~\bibnamefont
			{Mili{\'{c}}evi{\'{c}}}}, \bibinfo {author} {\bibfnamefont {T.}~\bibnamefont
			{Ozawa}}, \bibinfo {author} {\bibfnamefont {P.}~\bibnamefont {Andreakou}},
		\bibinfo {author} {\bibfnamefont {I.}~\bibnamefont {Carusotto}}, \bibinfo
		{author} {\bibfnamefont {T.}~\bibnamefont {Jacqmin}}, \bibinfo {author}
		{\bibfnamefont {E.}~\bibnamefont {Galopin}}, \bibinfo {author} {\bibfnamefont
			{A.}~\bibnamefont {Lema{\^{\i}}tre}}, \bibinfo {author} {\bibfnamefont
			{L.~L.}\ \bibnamefont {Gratiet}}, \bibinfo {author} {\bibfnamefont
			{I.}~\bibnamefont {Sagnes}}, \bibinfo {author} {\bibfnamefont
			{J.}~\bibnamefont {Bloch}},\ and\ \bibinfo {author} {\bibfnamefont
			{A.}~\bibnamefont {Amo}},\ }\bibfield  {title} {\bibinfo {title} {Edge states
			in polariton honeycomb lattices},\ }\href
	{https://doi.org/10.1088/2053-1583/2/3/034012} {\bibfield  {journal}
		{\bibinfo  {journal} {2D Mater.}\ }\textbf {\bibinfo {volume} {2}},\ \bibinfo
		{pages} {034012} (\bibinfo {year} {2015})}\BibitemShut {NoStop}%
	\bibitem [{\citenamefont {Gulevich}\ \emph {et~al.}(2016)\citenamefont
		{Gulevich}, \citenamefont {Yudin}, \citenamefont {Iorsh},\ and\ \citenamefont
		{Shelykh}}]{gulevich.prb.94.115437.2016}%
	\BibitemOpen
	\bibfield  {author} {\bibinfo {author} {\bibfnamefont {D.~R.}\ \bibnamefont
			{Gulevich}}, \bibinfo {author} {\bibfnamefont {D.}~\bibnamefont {Yudin}},
		\bibinfo {author} {\bibfnamefont {I.~V.}\ \bibnamefont {Iorsh}},\ and\
		\bibinfo {author} {\bibfnamefont {I.~A.}\ \bibnamefont {Shelykh}},\
	}\bibfield  {title} {\bibinfo {title} {Kagome lattice from exciton-polariton
			perspective},\ }\href {https://doi.org/10.1038/s41467-025-61120-6} {\bibfield
		{journal} {\bibinfo  {journal} {Phys. Rev. B}\ }\textbf {\bibinfo {volume}
			{94}},\ \bibinfo {pages} {115437} (\bibinfo {year} {2016})}\BibitemShut
	{NoStop}%
	\bibitem [{\citenamefont {Zhang}\ \emph
		{et~al.}(2018{\natexlab{a}})\citenamefont {Zhang}, \citenamefont {Kartashov},
		\citenamefont {Zhang}, \citenamefont {Torner},\ and\ \citenamefont
		{Skryabin}}]{zhang.lpr.12.1700348.2018}%
	\BibitemOpen
	\bibfield  {author} {\bibinfo {author} {\bibfnamefont {Y.~Q.}\ \bibnamefont
			{Zhang}}, \bibinfo {author} {\bibfnamefont {Y.~V.}\ \bibnamefont
			{Kartashov}}, \bibinfo {author} {\bibfnamefont {Y.~P.}\ \bibnamefont
			{Zhang}}, \bibinfo {author} {\bibfnamefont {L.}~\bibnamefont {Torner}},\ and\
		\bibinfo {author} {\bibfnamefont {D.~V.}\ \bibnamefont {Skryabin}},\
	}\bibfield  {title} {\bibinfo {title} {Resonant edge-state switching in
			polariton topological insulators},\ }\href
	{https://doi.org/https://doi.org/10.1002/lpor.201700348} {\bibfield
		{journal} {\bibinfo  {journal} {Laser Photon. Rev.}\ }\textbf {\bibinfo
			{volume} {12}},\ \bibinfo {pages} {1700348} (\bibinfo {year}
		{2018}{\natexlab{a}})}\BibitemShut {NoStop}%
	\bibitem [{\citenamefont {Zhang}\ \emph
		{et~al.}(2018{\natexlab{b}})\citenamefont {Zhang}, \citenamefont {Kartashov},
		\citenamefont {Zhang}, \citenamefont {Torner},\ and\ \citenamefont
		{Skryabin}}]{zhang.apl.3.120801.2018}%
	\BibitemOpen
	\bibfield  {author} {\bibinfo {author} {\bibfnamefont {Y.~Q.}\ \bibnamefont
			{Zhang}}, \bibinfo {author} {\bibfnamefont {Y.~V.}\ \bibnamefont
			{Kartashov}}, \bibinfo {author} {\bibfnamefont {Y.~P.}\ \bibnamefont
			{Zhang}}, \bibinfo {author} {\bibfnamefont {L.}~\bibnamefont {Torner}},\ and\
		\bibinfo {author} {\bibfnamefont {D.~V.}\ \bibnamefont {Skryabin}},\
	}\bibfield  {title} {\bibinfo {title} {Inhibition of tunneling and edge state
			control in polariton topological insulators},\ }\href
	{https://doi.org/10.1063/1.5043486} {\bibfield  {journal} {\bibinfo
			{journal} {APL Photon.}\ }\textbf {\bibinfo {volume} {3}},\ \bibinfo {pages}
		{120801} (\bibinfo {year} {2018}{\natexlab{b}})}\BibitemShut {NoStop}%
	\bibitem [{\citenamefont {Zhang}\ \emph
		{et~al.}(2019{\natexlab{a}})\citenamefont {Zhang}, \citenamefont {Chen},
		\citenamefont {Kartashov}, \citenamefont {Skryabin},\ and\ \citenamefont
		{Ye}}]{zhang.lpr.13.1900198.2019}%
	\BibitemOpen
	\bibfield  {author} {\bibinfo {author} {\bibfnamefont {W.}~\bibnamefont
			{Zhang}}, \bibinfo {author} {\bibfnamefont {X.}~\bibnamefont {Chen}},
		\bibinfo {author} {\bibfnamefont {Y.~V.}\ \bibnamefont {Kartashov}}, \bibinfo
		{author} {\bibfnamefont {D.~V.}\ \bibnamefont {Skryabin}},\ and\ \bibinfo
		{author} {\bibfnamefont {F.}~\bibnamefont {Ye}},\ }\bibfield  {title}
	{\bibinfo {title} {Finite-dimensional bistable topological insulators: From
			small to large},\ }\href {https://doi.org/10.1002/lpor.201900198} {\bibfield
		{journal} {\bibinfo  {journal} {Laser Photon. Rev.}\ }\textbf {\bibinfo
			{volume} {13}},\ \bibinfo {pages} {1900198} (\bibinfo {year}
		{2019}{\natexlab{a}})}\BibitemShut {NoStop}%
	\bibitem [{\citenamefont {Liu}\ \emph {et~al.}(2020{\natexlab{b}})\citenamefont
		{Liu}, \citenamefont {Ji}, \citenamefont {Wang}, \citenamefont {Modi},
		\citenamefont {Hwang}, \citenamefont {Zheng}, \citenamefont {Sorger},
		\citenamefont {Pan},\ and\ \citenamefont
		{Agarwal}}]{liu.science.370.600.2020}%
	\BibitemOpen
	\bibfield  {author} {\bibinfo {author} {\bibfnamefont {W.}~\bibnamefont
			{Liu}}, \bibinfo {author} {\bibfnamefont {Z.}~\bibnamefont {Ji}}, \bibinfo
		{author} {\bibfnamefont {Y.}~\bibnamefont {Wang}}, \bibinfo {author}
		{\bibfnamefont {G.}~\bibnamefont {Modi}}, \bibinfo {author} {\bibfnamefont
			{M.}~\bibnamefont {Hwang}}, \bibinfo {author} {\bibfnamefont
			{B.}~\bibnamefont {Zheng}}, \bibinfo {author} {\bibfnamefont {V.~J.}\
			\bibnamefont {Sorger}}, \bibinfo {author} {\bibfnamefont {A.}~\bibnamefont
			{Pan}},\ and\ \bibinfo {author} {\bibfnamefont {R.}~\bibnamefont {Agarwal}},\
	}\bibfield  {title} {\bibinfo {title} {Generation of helical topological
			exciton-polaritons},\ }\href {https://doi.org/10.1126/science.abc4975}
	{\bibfield  {journal} {\bibinfo  {journal} {Science}\ }\textbf {\bibinfo
			{volume} {370}},\ \bibinfo {pages} {600} (\bibinfo {year}
		{2020}{\natexlab{b}})}\BibitemShut {NoStop}%
	\bibitem [{\citenamefont {Peng}\ \emph {et~al.}(2024)\citenamefont {Peng},
		\citenamefont {Li}, \citenamefont {Sun}, \citenamefont {Rivero},
		\citenamefont {Ti}, \citenamefont {Han}, \citenamefont {Ge}, \citenamefont
		{Yang}, \citenamefont {Zhang},\ and\ \citenamefont {Bao}}]{peng.nn.2024}%
	\BibitemOpen
	\bibfield  {author} {\bibinfo {author} {\bibfnamefont {K.}~\bibnamefont
			{Peng}}, \bibinfo {author} {\bibfnamefont {W.}~\bibnamefont {Li}}, \bibinfo
		{author} {\bibfnamefont {M.}~\bibnamefont {Sun}}, \bibinfo {author}
		{\bibfnamefont {J.~D.~H.}\ \bibnamefont {Rivero}}, \bibinfo {author}
		{\bibfnamefont {C.}~\bibnamefont {Ti}}, \bibinfo {author} {\bibfnamefont
			{X.}~\bibnamefont {Han}}, \bibinfo {author} {\bibfnamefont {L.}~\bibnamefont
			{Ge}}, \bibinfo {author} {\bibfnamefont {L.}~\bibnamefont {Yang}}, \bibinfo
		{author} {\bibfnamefont {X.}~\bibnamefont {Zhang}},\ and\ \bibinfo {author}
		{\bibfnamefont {W.}~\bibnamefont {Bao}},\ }\bibfield  {title} {\bibinfo
		{title} {Topological valley {Hall} polariton condensation},\ }\href
	{https://doi.org/10.1038/s41565-024-01674-6} {\bibfield  {journal} {\bibinfo
			{journal} {Nat. Nanotech.}\ }\textbf {\bibinfo {volume} {19}},\ \bibinfo
		{pages} {1283} (\bibinfo {year} {2024})}\BibitemShut {NoStop}%
	\bibitem [{\citenamefont {Jin}\ \emph {et~al.}(2024)\citenamefont {Jin},
		\citenamefont {Mandal}, \citenamefont {Wu}, \citenamefont {Zhang},
		\citenamefont {Wen}, \citenamefont {Ren}, \citenamefont {Zhang},
		\citenamefont {Liew}, \citenamefont {Xiong},\ and\ \citenamefont
		{Su}}]{jin.nc.15.10563.2024}%
	\BibitemOpen
	\bibfield  {author} {\bibinfo {author} {\bibfnamefont {F.}~\bibnamefont
			{Jin}}, \bibinfo {author} {\bibfnamefont {S.}~\bibnamefont {Mandal}},
		\bibinfo {author} {\bibfnamefont {J.}~\bibnamefont {Wu}}, \bibinfo {author}
		{\bibfnamefont {Z.}~\bibnamefont {Zhang}}, \bibinfo {author} {\bibfnamefont
			{W.}~\bibnamefont {Wen}}, \bibinfo {author} {\bibfnamefont {J.}~\bibnamefont
			{Ren}}, \bibinfo {author} {\bibfnamefont {B.}~\bibnamefont {Zhang}}, \bibinfo
		{author} {\bibfnamefont {T.~C.~H.}\ \bibnamefont {Liew}}, \bibinfo {author}
		{\bibfnamefont {Q.}~\bibnamefont {Xiong}},\ and\ \bibinfo {author}
		{\bibfnamefont {R.}~\bibnamefont {Su}},\ }\bibfield  {title} {\bibinfo
		{title} {Observation of perovskite topological valley exciton-polaritons at
			room temperature},\ }\href {https://doi.org/10.1038/s41467-025-61120-6}
	{\bibfield  {journal} {\bibinfo  {journal} {Nat. Commun.}\ }\textbf {\bibinfo
			{volume} {15}},\ \bibinfo {pages} {10563} (\bibinfo {year}
		{2024})}\BibitemShut {NoStop}%
	\bibitem [{\citenamefont {Wu}\ \emph {et~al.}(2023)\citenamefont {Wu},
		\citenamefont {Ghosh}, \citenamefont {Gan}, \citenamefont {Shi},
		\citenamefont {Mandal}, \citenamefont {Sun}, \citenamefont {Zhang},
		\citenamefont {Liew}, \citenamefont {Su},\ and\ \citenamefont
		{Xiong}}]{wu.sa.9.eadg4322.2023}%
	\BibitemOpen
	\bibfield  {author} {\bibinfo {author} {\bibfnamefont {J.}~\bibnamefont
			{Wu}}, \bibinfo {author} {\bibfnamefont {S.}~\bibnamefont {Ghosh}}, \bibinfo
		{author} {\bibfnamefont {Y.}~\bibnamefont {Gan}}, \bibinfo {author}
		{\bibfnamefont {Y.}~\bibnamefont {Shi}}, \bibinfo {author} {\bibfnamefont
			{S.}~\bibnamefont {Mandal}}, \bibinfo {author} {\bibfnamefont
			{H.}~\bibnamefont {Sun}}, \bibinfo {author} {\bibfnamefont {B.}~\bibnamefont
			{Zhang}}, \bibinfo {author} {\bibfnamefont {T.~C.~H.}\ \bibnamefont {Liew}},
		\bibinfo {author} {\bibfnamefont {R.}~\bibnamefont {Su}},\ and\ \bibinfo
		{author} {\bibfnamefont {Q.}~\bibnamefont {Xiong}},\ }\bibfield  {title}
	{\bibinfo {title} {Higher-order topological polariton corner state lasing},\
	}\href {https://doi.org/10.1126/sciadv.adg4322} {\bibfield  {journal}
		{\bibinfo  {journal} {Sci. Adv.}\ }\textbf {\bibinfo {volume} {9}},\ \bibinfo
		{pages} {eadg4322} (\bibinfo {year} {2023})}\BibitemShut {NoStop}%
	\bibitem [{\citenamefont {Su}\ \emph {et~al.}(2021)\citenamefont {Su},
		\citenamefont {Ghosh}, \citenamefont {Liew},\ and\ \citenamefont
		{Xiong}}]{su.sa.7.8049.2021}%
	\BibitemOpen
	\bibfield  {author} {\bibinfo {author} {\bibfnamefont {R.}~\bibnamefont
			{Su}}, \bibinfo {author} {\bibfnamefont {S.}~\bibnamefont {Ghosh}}, \bibinfo
		{author} {\bibfnamefont {T.~C.~H.}\ \bibnamefont {Liew}},\ and\ \bibinfo
		{author} {\bibfnamefont {Q.}~\bibnamefont {Xiong}},\ }\bibfield  {title}
	{\bibinfo {title} {Optical switching of topological phase in a perovskite
			polariton lattice},\ }\href {https://doi.org/10.1038/s42254-025-00889-3}
	{\bibfield  {journal} {\bibinfo  {journal} {Sci. Adv.}\ }\textbf {\bibinfo
			{volume} {7}},\ \bibinfo {pages} {eabf8049} (\bibinfo {year}
		{2021})}\BibitemShut {NoStop}%
	\bibitem [{\citenamefont {Bennenhei}\ \emph {et~al.}(2024)\citenamefont
		{Bennenhei}, \citenamefont {Shan}, \citenamefont {Struve}, \citenamefont
		{Kunte}, \citenamefont {Eilenberger}, \citenamefont {Ohmer}, \citenamefont
		{Fischer}, \citenamefont {Schumacher}, \citenamefont {Ma}, \citenamefont
		{Schneider},\ and\ \citenamefont {Esmann}}]{bennenhei.acs.11.3046.2024}%
	\BibitemOpen
	\bibfield  {author} {\bibinfo {author} {\bibfnamefont {C.}~\bibnamefont
			{Bennenhei}}, \bibinfo {author} {\bibfnamefont {H.}~\bibnamefont {Shan}},
		\bibinfo {author} {\bibfnamefont {M.}~\bibnamefont {Struve}}, \bibinfo
		{author} {\bibfnamefont {N.}~\bibnamefont {Kunte}}, \bibinfo {author}
		{\bibfnamefont {F.}~\bibnamefont {Eilenberger}}, \bibinfo {author}
		{\bibfnamefont {J.}~\bibnamefont {Ohmer}}, \bibinfo {author} {\bibfnamefont
			{U.}~\bibnamefont {Fischer}}, \bibinfo {author} {\bibfnamefont
			{S.}~\bibnamefont {Schumacher}}, \bibinfo {author} {\bibfnamefont
			{X.}~\bibnamefont {Ma}}, \bibinfo {author} {\bibfnamefont {C.}~\bibnamefont
			{Schneider}},\ and\ \bibinfo {author} {\bibfnamefont {M.}~\bibnamefont
			{Esmann}},\ }\bibfield  {title} {\bibinfo {title} {Organic room-temperature
			polariton condensate in a higher-order topological lattice},\ }\href
	{https://doi.org/10.1038/s41467-025-61120-6} {\bibfield  {journal} {\bibinfo
			{journal} {ACS Photon.}\ }\textbf {\bibinfo {volume} {11}},\ \bibinfo {pages}
		{3046} (\bibinfo {year} {2024})}\BibitemShut {NoStop}%
	\bibitem [{\citenamefont {Jin}\ \emph {et~al.}(2025)\citenamefont {Jin},
		\citenamefont {Mandal}, \citenamefont {Wang}, \citenamefont {Zhang},\ and\
		\citenamefont {Su}}]{jin.nc.16.6002.2025}%
	\BibitemOpen
	\bibfield  {author} {\bibinfo {author} {\bibfnamefont {F.}~\bibnamefont
			{Jin}}, \bibinfo {author} {\bibfnamefont {S.}~\bibnamefont {Mandal}},
		\bibinfo {author} {\bibfnamefont {X.}~\bibnamefont {Wang}}, \bibinfo {author}
		{\bibfnamefont {B.}~\bibnamefont {Zhang}},\ and\ \bibinfo {author}
		{\bibfnamefont {R.}~\bibnamefont {Su}},\ }\bibfield  {title} {\bibinfo
		{title} {Perovskite topological exciton-polariton disclination laser at room
			temperature},\ }\href {https://doi.org/10.1038/s41467-025-61120-6} {\bibfield
		{journal} {\bibinfo  {journal} {Nat. Commun.}\ }\textbf {\bibinfo {volume}
			{16}},\ \bibinfo {pages} {6002} (\bibinfo {year} {2025})}\BibitemShut
	{NoStop}%
	\bibitem [{\citenamefont {St-Jean}\ \emph {et~al.}(2017)\citenamefont
		{St-Jean}, \citenamefont {Goblot}, \citenamefont {Galopin}, \citenamefont
		{Lema\^{\i}tre}, \citenamefont {Ozawa}, \citenamefont {Le~Gratiet},
		\citenamefont {Sagnes}, \citenamefont {Bloch},\ and\ \citenamefont
		{Amo}}]{jean.np.11.651.2017}%
	\BibitemOpen
	\bibfield  {author} {\bibinfo {author} {\bibfnamefont {P.}~\bibnamefont
			{St-Jean}}, \bibinfo {author} {\bibfnamefont {V.}~\bibnamefont {Goblot}},
		\bibinfo {author} {\bibfnamefont {E.}~\bibnamefont {Galopin}}, \bibinfo
		{author} {\bibfnamefont {A.}~\bibnamefont {Lema\^{\i}tre}}, \bibinfo {author}
		{\bibfnamefont {T.}~\bibnamefont {Ozawa}}, \bibinfo {author} {\bibfnamefont
			{L.}~\bibnamefont {Le~Gratiet}}, \bibinfo {author} {\bibfnamefont
			{I.}~\bibnamefont {Sagnes}}, \bibinfo {author} {\bibfnamefont
			{J.}~\bibnamefont {Bloch}},\ and\ \bibinfo {author} {\bibfnamefont
			{A.}~\bibnamefont {Amo}},\ }\bibfield  {title} {\bibinfo {title} {Lasing in
			topological edge states of a one-dimensional lattice},\ }\href
	{https://doi.org/10.1038/s41566-017-0006-2} {\bibfield  {journal} {\bibinfo
			{journal} {Nat. Photon.}\ }\textbf {\bibinfo {volume} {11}},\ \bibinfo
		{pages} {651} (\bibinfo {year} {2017})}\BibitemShut {NoStop}%
	\bibitem [{\citenamefont {del Valle Inclan~Redondo}\ \emph
		{et~al.}(2024)\citenamefont {del Valle Inclan~Redondo}, \citenamefont {Xu},
		\citenamefont {Liew}, \citenamefont {Ostrovskaya}, \citenamefont {Stegmaier},
		\citenamefont {Thomale}, \citenamefont {Schneider}, \citenamefont {Dam},
		\citenamefont {Klembt}, \citenamefont {H\"ofling}, \citenamefont {Tarucha},\
		and\ \citenamefont {Fraser}}]{redondo.np.18.548.2024}%
	\BibitemOpen
	\bibfield  {author} {\bibinfo {author} {\bibfnamefont {Y.}~\bibnamefont {del
				Valle Inclan~Redondo}}, \bibinfo {author} {\bibfnamefont {X.}~\bibnamefont
			{Xu}}, \bibinfo {author} {\bibfnamefont {T.~C.~H.}\ \bibnamefont {Liew}},
		\bibinfo {author} {\bibfnamefont {E.~A.}\ \bibnamefont {Ostrovskaya}},
		\bibinfo {author} {\bibfnamefont {A.}~\bibnamefont {Stegmaier}}, \bibinfo
		{author} {\bibfnamefont {R.}~\bibnamefont {Thomale}}, \bibinfo {author}
		{\bibfnamefont {C.}~\bibnamefont {Schneider}}, \bibinfo {author}
		{\bibfnamefont {S.}~\bibnamefont {Dam}}, \bibinfo {author} {\bibfnamefont
			{S.}~\bibnamefont {Klembt}}, \bibinfo {author} {\bibfnamefont
			{S.}~\bibnamefont {H\"ofling}}, \bibinfo {author} {\bibfnamefont
			{S.}~\bibnamefont {Tarucha}},\ and\ \bibinfo {author} {\bibfnamefont {M.~D.}\
			\bibnamefont {Fraser}},\ }\bibfield  {title} {\bibinfo {title}
		{Non-reciprocal band structures in an exciton-polariton {Floquet} optical
			lattice},\ }\href {https://doi.org/10.1038/s41566-024-01424-z} {\bibfield
		{journal} {\bibinfo  {journal} {Nat. Photon.}\ }\textbf {\bibinfo {volume}
			{18}},\ \bibinfo {pages} {548} (\bibinfo {year} {2024})}\BibitemShut
	{NoStop}%
	\bibitem [{\citenamefont {Liang}\ \emph {et~al.}(2026)\citenamefont {Liang},
		\citenamefont {Zheng}, \citenamefont {Jin}, \citenamefont {Bao},
		\citenamefont {Dini}, \citenamefont {Ren}, \citenamefont {Liu}, \citenamefont
		{Kr\'ol}, \citenamefont {Ostrovskaya}, \citenamefont {Estrecho},
		\citenamefont {Zhang}, \citenamefont {Liew},\ and\ \citenamefont
		{Su}}]{liang.np.22.151.2026}%
	\BibitemOpen
	\bibfield  {author} {\bibinfo {author} {\bibfnamefont {J.}~\bibnamefont
			{Liang}}, \bibinfo {author} {\bibfnamefont {H.}~\bibnamefont {Zheng}},
		\bibinfo {author} {\bibfnamefont {F.}~\bibnamefont {Jin}}, \bibinfo {author}
		{\bibfnamefont {R.}~\bibnamefont {Bao}}, \bibinfo {author} {\bibfnamefont
			{K.}~\bibnamefont {Dini}}, \bibinfo {author} {\bibfnamefont {J.}~\bibnamefont
			{Ren}}, \bibinfo {author} {\bibfnamefont {Y.}~\bibnamefont {Liu}}, \bibinfo
		{author} {\bibfnamefont {M.}~\bibnamefont {Kr\'ol}}, \bibinfo {author}
		{\bibfnamefont {E.~A.}\ \bibnamefont {Ostrovskaya}}, \bibinfo {author}
		{\bibfnamefont {E.}~\bibnamefont {Estrecho}}, \bibinfo {author}
		{\bibfnamefont {B.}~\bibnamefont {Zhang}}, \bibinfo {author} {\bibfnamefont
			{T.~C.~H.}\ \bibnamefont {Liew}},\ and\ \bibinfo {author} {\bibfnamefont
			{R.}~\bibnamefont {Su}},\ }\bibfield  {title} {\bibinfo {title}
		{Twist-induced non-{Hermitian} topology of exciton-polaritons},\ }\href
	{https://doi.org/10.1038/s41567-025-03115-0} {\bibfield  {journal} {\bibinfo
			{journal} {Nat. Phys.}\ }\textbf {\bibinfo {volume} {22}},\ \bibinfo {pages}
		{151} (\bibinfo {year} {2026})}\BibitemShut {NoStop}%
	\bibitem [{\citenamefont {Solnyshkov}\ \emph {et~al.}(2021)\citenamefont
		{Solnyshkov}, \citenamefont {Malpuech}, \citenamefont {St-Jean},
		\citenamefont {Ravets}, \citenamefont {Bloch},\ and\ \citenamefont
		{Amo}}]{solnyshkov.ome.11.1119.2021}%
	\BibitemOpen
	\bibfield  {author} {\bibinfo {author} {\bibfnamefont {D.~D.}\ \bibnamefont
			{Solnyshkov}}, \bibinfo {author} {\bibfnamefont {G.}~\bibnamefont
			{Malpuech}}, \bibinfo {author} {\bibfnamefont {P.}~\bibnamefont {St-Jean}},
		\bibinfo {author} {\bibfnamefont {S.}~\bibnamefont {Ravets}}, \bibinfo
		{author} {\bibfnamefont {J.}~\bibnamefont {Bloch}},\ and\ \bibinfo {author}
		{\bibfnamefont {A.}~\bibnamefont {Amo}},\ }\bibfield  {title} {\bibinfo
		{title} {Microcavity polaritons for topological photonics [{Invited}]},\
	}\href {https://doi.org/10.1364/OME.414890} {\bibfield  {journal} {\bibinfo
			{journal} {Opt. Mater. Express}\ }\textbf {\bibinfo {volume} {11}},\ \bibinfo
		{pages} {1119} (\bibinfo {year} {2021})}\BibitemShut {NoStop}%
	\bibitem [{\citenamefont {Luo}\ \emph {et~al.}(2023)\citenamefont {Luo},
		\citenamefont {Zhou}, \citenamefont {Zhang},\ and\ \citenamefont
		{Chen}}]{luo.apr.10.011316.2023}%
	\BibitemOpen
	\bibfield  {author} {\bibinfo {author} {\bibfnamefont {S.}~\bibnamefont
			{Luo}}, \bibinfo {author} {\bibfnamefont {H.}~\bibnamefont {Zhou}}, \bibinfo
		{author} {\bibfnamefont {L.}~\bibnamefont {Zhang}},\ and\ \bibinfo {author}
		{\bibfnamefont {Z.}~\bibnamefont {Chen}},\ }\bibfield  {title} {\bibinfo
		{title} {Nanophotonics of microcavity exciton–polaritons},\ }\href
	{https://doi.org/10.1063/5.0121316} {\bibfield  {journal} {\bibinfo
			{journal} {Appl. Phys. Rev.}\ }\textbf {\bibinfo {volume} {10}},\ \bibinfo
		{pages} {011316} (\bibinfo {year} {2023})}\BibitemShut {NoStop}%
	\bibitem [{\citenamefont {Kartashov}\ and\ \citenamefont
		{Skryabin}(2016)}]{kartashov.optica.3.1228.2016}%
	\BibitemOpen
	\bibfield  {author} {\bibinfo {author} {\bibfnamefont {Y.~V.}\ \bibnamefont
			{Kartashov}}\ and\ \bibinfo {author} {\bibfnamefont {D.~V.}\ \bibnamefont
			{Skryabin}},\ }\bibfield  {title} {\bibinfo {title} {Modulational instability
			and solitary waves in polariton topological insulators},\ }\href
	{https://doi.org/10.1364/OPTICA.3.001228} {\bibfield  {journal} {\bibinfo
			{journal} {Optica}\ }\textbf {\bibinfo {volume} {3}},\ \bibinfo {pages}
		{1228} (\bibinfo {year} {2016})}\BibitemShut {NoStop}%
	\bibitem [{\citenamefont {Gulevich}\ \emph {et~al.}(2017)\citenamefont
		{Gulevich}, \citenamefont {Yudin}, \citenamefont {Skryabin}, \citenamefont
		{Iorsh},\ and\ \citenamefont {Shelykh}}]{gulevich.sr.7.1780.2017}%
	\BibitemOpen
	\bibfield  {author} {\bibinfo {author} {\bibfnamefont {D.~R.}\ \bibnamefont
			{Gulevich}}, \bibinfo {author} {\bibfnamefont {D.}~\bibnamefont {Yudin}},
		\bibinfo {author} {\bibfnamefont {D.~V.}\ \bibnamefont {Skryabin}}, \bibinfo
		{author} {\bibfnamefont {I.~V.}\ \bibnamefont {Iorsh}},\ and\ \bibinfo
		{author} {\bibfnamefont {I.~A.}\ \bibnamefont {Shelykh}},\ }\bibfield
	{title} {\bibinfo {title} {Exploring nonlinear topological states of matter
			with exciton-polaritons: Edge solitons in kagome lattice},\ }\href
	{https://doi.org/10.1038/s41598-017-01646-y} {\bibfield  {journal} {\bibinfo
			{journal} {Sci. Rep.}\ }\textbf {\bibinfo {volume} {7}},\ \bibinfo {pages}
		{1780} (\bibinfo {year} {2017})}\BibitemShut {NoStop}%
	\bibitem [{\citenamefont {Bleu}\ \emph {et~al.}(2016)\citenamefont {Bleu},
		\citenamefont {Solnyshkov},\ and\ \citenamefont
		{Malpuech}}]{bleu.prb.93.085438.2016}%
	\BibitemOpen
	\bibfield  {author} {\bibinfo {author} {\bibfnamefont {O.}~\bibnamefont
			{Bleu}}, \bibinfo {author} {\bibfnamefont {D.~D.}\ \bibnamefont
			{Solnyshkov}},\ and\ \bibinfo {author} {\bibfnamefont {G.}~\bibnamefont
			{Malpuech}},\ }\bibfield  {title} {\bibinfo {title} {Interacting quantum
			fluid in a polariton {C}hern insulator},\ }\href
	{https://doi.org/10.1103/PhysRevB.93.085438} {\bibfield  {journal} {\bibinfo
			{journal} {Phys. Rev. B}\ }\textbf {\bibinfo {volume} {93}},\ \bibinfo
		{pages} {085438} (\bibinfo {year} {2016})}\BibitemShut {NoStop}%
	\bibitem [{\citenamefont {Kartashov}\ and\ \citenamefont
		{Skryabin}(2017)}]{kartashov.prl.119.253904.2017}%
	\BibitemOpen
	\bibfield  {author} {\bibinfo {author} {\bibfnamefont {Y.~V.}\ \bibnamefont
			{Kartashov}}\ and\ \bibinfo {author} {\bibfnamefont {D.~V.}\ \bibnamefont
			{Skryabin}},\ }\bibfield  {title} {\bibinfo {title} {Bistable topological
			insulator with exciton-polaritons},\ }\href
	{https://doi.org/10.1103/PhysRevLett.119.253904} {\bibfield  {journal}
		{\bibinfo  {journal} {Phys. Rev. Lett.}\ }\textbf {\bibinfo {volume} {119}},\
		\bibinfo {pages} {253904} (\bibinfo {year} {2017})}\BibitemShut {NoStop}%
	\bibitem [{\citenamefont {Li}\ \emph {et~al.}(2018)\citenamefont {Li},
		\citenamefont {Ye}, \citenamefont {Chen}, \citenamefont {Kartashov},
		\citenamefont {Ferrando}, \citenamefont {Torner},\ and\ \citenamefont
		{Skryabin}}]{li.prb.97.081103.2018}%
	\BibitemOpen
	\bibfield  {author} {\bibinfo {author} {\bibfnamefont {C.}~\bibnamefont
			{Li}}, \bibinfo {author} {\bibfnamefont {F.}~\bibnamefont {Ye}}, \bibinfo
		{author} {\bibfnamefont {X.}~\bibnamefont {Chen}}, \bibinfo {author}
		{\bibfnamefont {Y.~V.}\ \bibnamefont {Kartashov}}, \bibinfo {author}
		{\bibfnamefont {A.}~\bibnamefont {Ferrando}}, \bibinfo {author}
		{\bibfnamefont {L.}~\bibnamefont {Torner}},\ and\ \bibinfo {author}
		{\bibfnamefont {D.~V.}\ \bibnamefont {Skryabin}},\ }\bibfield  {title}
	{\bibinfo {title} {Lieb polariton topological insulators},\ }\href
	{https://doi.org/10.1103/PhysRevB.97.081103} {\bibfield  {journal} {\bibinfo
			{journal} {Phys. Rev. B}\ }\textbf {\bibinfo {volume} {97}},\ \bibinfo
		{pages} {081103} (\bibinfo {year} {2018})}\BibitemShut {NoStop}%
	\bibitem [{\citenamefont {Bleu}\ \emph {et~al.}(2018)\citenamefont {Bleu},
		\citenamefont {Malpuech},\ and\ \citenamefont
		{Solnyshkov}}]{bleu.nc.9.3991.2018}%
	\BibitemOpen
	\bibfield  {author} {\bibinfo {author} {\bibfnamefont {O.}~\bibnamefont
			{Bleu}}, \bibinfo {author} {\bibfnamefont {G.}~\bibnamefont {Malpuech}},\
		and\ \bibinfo {author} {\bibfnamefont {D.~D.}\ \bibnamefont {Solnyshkov}},\
	}\bibfield  {title} {\bibinfo {title} {Robust quantum valley {H}all effect
			for vortices in an interacting bosonic quantum fluid},\ }\href
	{https://doi.org/10.1038/s41467-018-06520-7} {\bibfield  {journal} {\bibinfo
			{journal} {Nat. Commun.}\ }\textbf {\bibinfo {volume} {9}},\ \bibinfo {pages}
		{3991} (\bibinfo {year} {2018})}\BibitemShut {NoStop}%
	\bibitem [{\citenamefont {Kartashov}\ and\ \citenamefont
		{Skryabin}(2019)}]{kartashov.prl.122.083902.2019}%
	\BibitemOpen
	\bibfield  {author} {\bibinfo {author} {\bibfnamefont {Y.~V.}\ \bibnamefont
			{Kartashov}}\ and\ \bibinfo {author} {\bibfnamefont {D.~V.}\ \bibnamefont
			{Skryabin}},\ }\bibfield  {title} {\bibinfo {title} {Two-dimensional
			topological polariton laser},\ }\href
	{https://doi.org/10.1103/PhysRevLett.122.083902} {\bibfield  {journal}
		{\bibinfo  {journal} {Phys. Rev. Lett.}\ }\textbf {\bibinfo {volume} {122}},\
		\bibinfo {pages} {083902} (\bibinfo {year} {2019})}\BibitemShut {NoStop}%
	\bibitem [{\citenamefont {Zhang}\ \emph
		{et~al.}(2019{\natexlab{b}})\citenamefont {Zhang}, \citenamefont
		{Kartashov},\ and\ \citenamefont {Ferrando}}]{zhang.pra.99.053836.2019}%
	\BibitemOpen
	\bibfield  {author} {\bibinfo {author} {\bibfnamefont {Y.~Q.}\ \bibnamefont
			{Zhang}}, \bibinfo {author} {\bibfnamefont {Y.~V.}\ \bibnamefont
			{Kartashov}},\ and\ \bibinfo {author} {\bibfnamefont {A.}~\bibnamefont
			{Ferrando}},\ }\bibfield  {title} {\bibinfo {title} {Interface states in
			polariton topological insulators},\ }\href
	{https://doi.org/10.1103/PhysRevA.99.053836} {\bibfield  {journal} {\bibinfo
			{journal} {Phys. Rev. A}\ }\textbf {\bibinfo {volume} {99}},\ \bibinfo
		{pages} {053836} (\bibinfo {year} {2019}{\natexlab{b}})}\BibitemShut
	{NoStop}%
	\bibitem [{\citenamefont {Smirnova}\ \emph {et~al.}(2020)\citenamefont
		{Smirnova}, \citenamefont {Leykam}, \citenamefont {Chong},\ and\
		\citenamefont {Kivshar}}]{smirnova.apr.7.021306.2020}%
	\BibitemOpen
	\bibfield  {author} {\bibinfo {author} {\bibfnamefont {D.}~\bibnamefont
			{Smirnova}}, \bibinfo {author} {\bibfnamefont {D.}~\bibnamefont {Leykam}},
		\bibinfo {author} {\bibfnamefont {Y.}~\bibnamefont {Chong}},\ and\ \bibinfo
		{author} {\bibfnamefont {Y.}~\bibnamefont {Kivshar}},\ }\bibfield  {title}
	{\bibinfo {title} {Nonlinear topological photonics},\ }\href
	{https://doi.org/10.1063/1.5142397} {\bibfield  {journal} {\bibinfo
			{journal} {Appl. Phys. Rev.}\ }\textbf {\bibinfo {volume} {7}},\ \bibinfo
		{pages} {021306} (\bibinfo {year} {2020})}\BibitemShut {NoStop}%
	\bibitem [{\citenamefont {Szameit}\ and\ \citenamefont
		{Rechtsman}(2024)}]{szameit.np.20.905.2024}%
	\BibitemOpen
	\bibfield  {author} {\bibinfo {author} {\bibfnamefont {A.}~\bibnamefont
			{Szameit}}\ and\ \bibinfo {author} {\bibfnamefont {M.~C.}\ \bibnamefont
			{Rechtsman}},\ }\bibfield  {title} {\bibinfo {title} {Discrete nonlinear
			topological photonics},\ }\href {https://doi.org/10.1038/s41567-024-02454-8}
	{\bibfield  {journal} {\bibinfo  {journal} {Nat. Phys.}\ }\textbf {\bibinfo
			{volume} {20}},\ \bibinfo {pages} {905} (\bibinfo {year} {2024})}\BibitemShut
	{NoStop}%
	\bibitem [{\citenamefont {Lumer}\ \emph {et~al.}(2013)\citenamefont {Lumer},
		\citenamefont {Plotnik}, \citenamefont {Rechtsman},\ and\ \citenamefont
		{Segev}}]{lumer.prl.111.243905.2013}%
	\BibitemOpen
	\bibfield  {author} {\bibinfo {author} {\bibfnamefont {Y.}~\bibnamefont
			{Lumer}}, \bibinfo {author} {\bibfnamefont {Y.}~\bibnamefont {Plotnik}},
		\bibinfo {author} {\bibfnamefont {M.~C.}\ \bibnamefont {Rechtsman}},\ and\
		\bibinfo {author} {\bibfnamefont {M.}~\bibnamefont {Segev}},\ }\bibfield
	{title} {\bibinfo {title} {Self-localized states in photonic topological
			insulators},\ }\href {https://doi.org/10.1103/PhysRevLett.111.243905}
	{\bibfield  {journal} {\bibinfo  {journal} {Phys. Rev. Lett.}\ }\textbf
		{\bibinfo {volume} {111}},\ \bibinfo {pages} {243905} (\bibinfo {year}
		{2013})}\BibitemShut {NoStop}%
	\bibitem [{\citenamefont {Leykam}\ and\ \citenamefont
		{Chong}(2016)}]{leykam.prl.117.143901.2016}%
	\BibitemOpen
	\bibfield  {author} {\bibinfo {author} {\bibfnamefont {D.}~\bibnamefont
			{Leykam}}\ and\ \bibinfo {author} {\bibfnamefont {Y.~D.}\ \bibnamefont
			{Chong}},\ }\bibfield  {title} {\bibinfo {title} {Edge solitons in
			nonlinear-photonic topological insulators},\ }\href
	{https://doi.org/10.1103/PhysRevLett.117.143901} {\bibfield  {journal}
		{\bibinfo  {journal} {Phys. Rev. Lett.}\ }\textbf {\bibinfo {volume} {117}},\
		\bibinfo {pages} {143901} (\bibinfo {year} {2016})}\BibitemShut {NoStop}%
	\bibitem [{\citenamefont {Zhang}\ \emph
		{et~al.}(2019{\natexlab{c}})\citenamefont {Zhang}, \citenamefont {Chen},
		\citenamefont {Kartashov}, \citenamefont {Konotop},\ and\ \citenamefont
		{Ye}}]{zhang.prl.123.254103.2019}%
	\BibitemOpen
	\bibfield  {author} {\bibinfo {author} {\bibfnamefont {W.}~\bibnamefont
			{Zhang}}, \bibinfo {author} {\bibfnamefont {X.}~\bibnamefont {Chen}},
		\bibinfo {author} {\bibfnamefont {Y.~V.}\ \bibnamefont {Kartashov}}, \bibinfo
		{author} {\bibfnamefont {V.~V.}\ \bibnamefont {Konotop}},\ and\ \bibinfo
		{author} {\bibfnamefont {F.}~\bibnamefont {Ye}},\ }\bibfield  {title}
	{\bibinfo {title} {Coupling of edge states and topological {B}ragg
			solitons},\ }\href {https://doi.org/10.1103/PhysRevLett.123.254103}
	{\bibfield  {journal} {\bibinfo  {journal} {Phys. Rev. Lett.}\ }\textbf
		{\bibinfo {volume} {123}},\ \bibinfo {pages} {254103} (\bibinfo {year}
		{2019}{\natexlab{c}})}\BibitemShut {NoStop}%
	\bibitem [{\citenamefont {Ivanov}\ \emph {et~al.}(2020)\citenamefont {Ivanov},
		\citenamefont {Kartashov}, \citenamefont {Szameit}, \citenamefont {Torner},\
		and\ \citenamefont {Konotop}}]{ivanov.acs.7.735.2020}%
	\BibitemOpen
	\bibfield  {author} {\bibinfo {author} {\bibfnamefont {S.~K.}\ \bibnamefont
			{Ivanov}}, \bibinfo {author} {\bibfnamefont {Y.~V.}\ \bibnamefont
			{Kartashov}}, \bibinfo {author} {\bibfnamefont {A.}~\bibnamefont {Szameit}},
		\bibinfo {author} {\bibfnamefont {L.}~\bibnamefont {Torner}},\ and\ \bibinfo
		{author} {\bibfnamefont {V.~V.}\ \bibnamefont {Konotop}},\ }\bibfield
	{title} {\bibinfo {title} {Vector topological edge solitons in {F}loquet
			insulators},\ }\href {https://doi.org/10.1021/acsphotonics.9b01589}
	{\bibfield  {journal} {\bibinfo  {journal} {ACS Photon.}\ }\textbf {\bibinfo
			{volume} {7}},\ \bibinfo {pages} {735} (\bibinfo {year} {2020})}\BibitemShut
	{NoStop}%
	\bibitem [{\citenamefont {Mukherjee}\ and\ \citenamefont
		{Rechtsman}(2020)}]{mukherjee.science.368.856.2020}%
	\BibitemOpen
	\bibfield  {author} {\bibinfo {author} {\bibfnamefont {S.}~\bibnamefont
			{Mukherjee}}\ and\ \bibinfo {author} {\bibfnamefont {M.~C.}\ \bibnamefont
			{Rechtsman}},\ }\bibfield  {title} {\bibinfo {title} {Observation of
			{F}loquet solitons in a topological bandgap},\ }\href
	{https://doi.org/10.1126/science.aba8725} {\bibfield  {journal} {\bibinfo
			{journal} {Science}\ }\textbf {\bibinfo {volume} {368}},\ \bibinfo {pages}
		{856} (\bibinfo {year} {2020})}\BibitemShut {NoStop}%
	\bibitem [{\citenamefont {Maczewsky}\ \emph {et~al.}(2020)\citenamefont
		{Maczewsky}, \citenamefont {Heinrich}, \citenamefont {Kremer}, \citenamefont
		{Ivanov}, \citenamefont {Ehrhardt}, \citenamefont {Martinez}, \citenamefont
		{Kartashov}, \citenamefont {Konotop}, \citenamefont {Torner}, \citenamefont
		{Bauer},\ and\ \citenamefont {Szameit}}]{maczewsky.science.370.701.2020}%
	\BibitemOpen
	\bibfield  {author} {\bibinfo {author} {\bibfnamefont {L.~J.}\ \bibnamefont
			{Maczewsky}}, \bibinfo {author} {\bibfnamefont {M.}~\bibnamefont {Heinrich}},
		\bibinfo {author} {\bibfnamefont {M.}~\bibnamefont {Kremer}}, \bibinfo
		{author} {\bibfnamefont {S.~K.}\ \bibnamefont {Ivanov}}, \bibinfo {author}
		{\bibfnamefont {M.}~\bibnamefont {Ehrhardt}}, \bibinfo {author}
		{\bibfnamefont {F.}~\bibnamefont {Martinez}}, \bibinfo {author}
		{\bibfnamefont {Y.~V.}\ \bibnamefont {Kartashov}}, \bibinfo {author}
		{\bibfnamefont {V.~V.}\ \bibnamefont {Konotop}}, \bibinfo {author}
		{\bibfnamefont {L.}~\bibnamefont {Torner}}, \bibinfo {author} {\bibfnamefont
			{D.}~\bibnamefont {Bauer}},\ and\ \bibinfo {author} {\bibfnamefont
			{A.}~\bibnamefont {Szameit}},\ }\bibfield  {title} {\bibinfo {title}
		{Nonlinearity-induced photonic topological insulator},\ }\href
	{https://doi.org/10.1126/science.abd2033} {\bibfield  {journal} {\bibinfo
			{journal} {Science}\ }\textbf {\bibinfo {volume} {370}},\ \bibinfo {pages}
		{701} (\bibinfo {year} {2020})}\BibitemShut {NoStop}%
	\bibitem [{\citenamefont {Mukherjee}\ and\ \citenamefont
		{Rechtsman}(2021)}]{mukherjee.prx.11.041057.2021}%
	\BibitemOpen
	\bibfield  {author} {\bibinfo {author} {\bibfnamefont {S.}~\bibnamefont
			{Mukherjee}}\ and\ \bibinfo {author} {\bibfnamefont {M.~C.}\ \bibnamefont
			{Rechtsman}},\ }\bibfield  {title} {\bibinfo {title} {Observation of
			unidirectional solitonlike edge states in nonlinear {Floquet} topological
			insulators},\ }\href {https://doi.org/10.1103/PhysRevX.11.041057} {\bibfield
		{journal} {\bibinfo  {journal} {Phys. Rev. X}\ }\textbf {\bibinfo {volume}
			{11}},\ \bibinfo {pages} {041057} (\bibinfo {year} {2021})}\BibitemShut
	{NoStop}%
	\bibitem [{\citenamefont {Zhong}\ \emph {et~al.}(2021)\citenamefont {Zhong},
		\citenamefont {Xia}, \citenamefont {Zhang}, \citenamefont {Li}, \citenamefont
		{Song}, \citenamefont {Liu},\ and\ \citenamefont
		{Chen}}]{zhong.ap.3.056001.2021}%
	\BibitemOpen
	\bibfield  {author} {\bibinfo {author} {\bibfnamefont {H.}~\bibnamefont
			{Zhong}}, \bibinfo {author} {\bibfnamefont {S.}~\bibnamefont {Xia}}, \bibinfo
		{author} {\bibfnamefont {Y.}~\bibnamefont {Zhang}}, \bibinfo {author}
		{\bibfnamefont {Y.}~\bibnamefont {Li}}, \bibinfo {author} {\bibfnamefont
			{D.}~\bibnamefont {Song}}, \bibinfo {author} {\bibfnamefont {C.}~\bibnamefont
			{Liu}},\ and\ \bibinfo {author} {\bibfnamefont {Z.}~\bibnamefont {Chen}},\
	}\bibfield  {title} {\bibinfo {title} {{Nonlinear topological valley {Hall}
				edge states arising from type-{II} {Dirac} cones}},\ }\href
	{https://doi.org/10.1117/1.AP.3.5.056001} {\bibfield  {journal} {\bibinfo
			{journal} {Adv. Photon.}\ }\textbf {\bibinfo {volume} {3}},\ \bibinfo {pages}
		{056001} (\bibinfo {year} {2021})}\BibitemShut {NoStop}%
	\bibitem [{\citenamefont {Tang}\ \emph {et~al.}(2021)\citenamefont {Tang},
		\citenamefont {Ren}, \citenamefont {Kompanets}, \citenamefont {Kartashov},
		\citenamefont {Li},\ and\ \citenamefont {Zhang}}]{tang.oe.29.39755.2021}%
	\BibitemOpen
	\bibfield  {author} {\bibinfo {author} {\bibfnamefont {Q.}~\bibnamefont
			{Tang}}, \bibinfo {author} {\bibfnamefont {B.}~\bibnamefont {Ren}}, \bibinfo
		{author} {\bibfnamefont {V.~O.}\ \bibnamefont {Kompanets}}, \bibinfo {author}
		{\bibfnamefont {Y.~V.}\ \bibnamefont {Kartashov}}, \bibinfo {author}
		{\bibfnamefont {Y.}~\bibnamefont {Li}},\ and\ \bibinfo {author}
		{\bibfnamefont {Y.}~\bibnamefont {Zhang}},\ }\bibfield  {title} {\bibinfo
		{title} {Valley {Hall} edge solitons in a photonic graphene},\ }\href
	{https://doi.org/10.1364/OE.442338} {\bibfield  {journal} {\bibinfo
			{journal} {Opt. Express}\ }\textbf {\bibinfo {volume} {29}},\ \bibinfo
		{pages} {39755} (\bibinfo {year} {2021})}\BibitemShut {NoStop}%
	\bibitem [{\citenamefont {Ren}\ \emph {et~al.}(2021)\citenamefont {Ren},
		\citenamefont {Wang}, \citenamefont {Kompanets}, \citenamefont {Kartashov},
		\citenamefont {Li},\ and\ \citenamefont {Zhang}}]{ren.nano.10.3559.2021}%
	\BibitemOpen
	\bibfield  {author} {\bibinfo {author} {\bibfnamefont {B.}~\bibnamefont
			{Ren}}, \bibinfo {author} {\bibfnamefont {H.}~\bibnamefont {Wang}}, \bibinfo
		{author} {\bibfnamefont {V.~O.}\ \bibnamefont {Kompanets}}, \bibinfo {author}
		{\bibfnamefont {Y.~V.}\ \bibnamefont {Kartashov}}, \bibinfo {author}
		{\bibfnamefont {Y.}~\bibnamefont {Li}},\ and\ \bibinfo {author}
		{\bibfnamefont {Y.}~\bibnamefont {Zhang}},\ }\bibfield  {title} {\bibinfo
		{title} {Dark topological valley {Hall} edge solitons},\ }\href
	{https://doi.org/doi:10.1515/nanoph-2021-0385} {\bibfield  {journal}
		{\bibinfo  {journal} {Nanophoton.}\ }\textbf {\bibinfo {volume} {10}},\
		\bibinfo {pages} {3559} (\bibinfo {year} {2021})}\BibitemShut {NoStop}%
	\bibitem [{\citenamefont {Tang}\ \emph
		{et~al.}(2022{\natexlab{b}})\citenamefont {Tang}, \citenamefont {Ren},
		\citenamefont {Beli\'c}, \citenamefont {Zhang},\ and\ \citenamefont
		{Li}}]{tang.rrp.74.504.2022}%
	\BibitemOpen
	\bibfield  {author} {\bibinfo {author} {\bibfnamefont {Q.}~\bibnamefont
			{Tang}}, \bibinfo {author} {\bibfnamefont {B.}~\bibnamefont {Ren}}, \bibinfo
		{author} {\bibfnamefont {M.~R.}\ \bibnamefont {Beli\'c}}, \bibinfo {author}
		{\bibfnamefont {Y.}~\bibnamefont {Zhang}},\ and\ \bibinfo {author}
		{\bibfnamefont {Y.}~\bibnamefont {Li}},\ }\bibfield  {title} {\bibinfo
		{title} {Valley {Hall} edge solitons in the kagome photonic lattice},\
	}\href@noop {} {\bibfield  {journal} {\bibinfo  {journal} {Rom. Rep. Phys.}\
		}\textbf {\bibinfo {volume} {74}},\ \bibinfo {pages} {504} (\bibinfo {year}
		{2022}{\natexlab{b}})}\BibitemShut {NoStop}%
	\bibitem [{\citenamefont {Sala}\ \emph {et~al.}(2015)\citenamefont {Sala},
		\citenamefont {Solnyshkov}, \citenamefont {Carusotto}, \citenamefont
		{Jacqmin}, \citenamefont {Lema\^{\i}tre}, \citenamefont
		{Ter\ifmmode~\mbox{\c{c}}\else \c{c}\fi{}as}, \citenamefont {Nalitov},
		\citenamefont {Abbarchi}, \citenamefont {Galopin}, \citenamefont {Sagnes},
		\citenamefont {Bloch}, \citenamefont {Malpuech},\ and\ \citenamefont
		{Amo}}]{sala.prx.5.011034.2015}%
	\BibitemOpen
	\bibfield  {author} {\bibinfo {author} {\bibfnamefont {V.~G.}\ \bibnamefont
			{Sala}}, \bibinfo {author} {\bibfnamefont {D.~D.}\ \bibnamefont
			{Solnyshkov}}, \bibinfo {author} {\bibfnamefont {I.}~\bibnamefont
			{Carusotto}}, \bibinfo {author} {\bibfnamefont {T.}~\bibnamefont {Jacqmin}},
		\bibinfo {author} {\bibfnamefont {A.}~\bibnamefont {Lema\^{\i}tre}}, \bibinfo
		{author} {\bibfnamefont {H.}~\bibnamefont {Ter\ifmmode~\mbox{\c{c}}\else
				\c{c}\fi{}as}}, \bibinfo {author} {\bibfnamefont {A.}~\bibnamefont
			{Nalitov}}, \bibinfo {author} {\bibfnamefont {M.}~\bibnamefont {Abbarchi}},
		\bibinfo {author} {\bibfnamefont {E.}~\bibnamefont {Galopin}}, \bibinfo
		{author} {\bibfnamefont {I.}~\bibnamefont {Sagnes}}, \bibinfo {author}
		{\bibfnamefont {J.}~\bibnamefont {Bloch}}, \bibinfo {author} {\bibfnamefont
			{G.}~\bibnamefont {Malpuech}},\ and\ \bibinfo {author} {\bibfnamefont
			{A.}~\bibnamefont {Amo}},\ }\bibfield  {title} {\bibinfo {title} {Spin-orbit
			coupling for photons and polaritons in microstructures},\ }\href
	{https://doi.org/10.1103/PhysRevX.5.011034} {\bibfield  {journal} {\bibinfo
			{journal} {Phys. Rev. X}\ }\textbf {\bibinfo {volume} {5}},\ \bibinfo {pages}
		{011034} (\bibinfo {year} {2015})}\BibitemShut {NoStop}%
	\bibitem [{\citenamefont {Dufferwiel}\ \emph {et~al.}(2015)\citenamefont
		{Dufferwiel}, \citenamefont {Li}, \citenamefont {Cancellieri}, \citenamefont
		{Giriunas}, \citenamefont {Trichet}, \citenamefont {Whittaker}, \citenamefont
		{Walker}, \citenamefont {Fras}, \citenamefont {Clarke}, \citenamefont
		{Smith}, \citenamefont {Skolnick},\ and\ \citenamefont
		{Krizhanovskii}}]{dufferwiel.prl.115.246401.2015}%
	\BibitemOpen
	\bibfield  {author} {\bibinfo {author} {\bibfnamefont {S.}~\bibnamefont
			{Dufferwiel}}, \bibinfo {author} {\bibfnamefont {F.}~\bibnamefont {Li}},
		\bibinfo {author} {\bibfnamefont {E.}~\bibnamefont {Cancellieri}}, \bibinfo
		{author} {\bibfnamefont {L.}~\bibnamefont {Giriunas}}, \bibinfo {author}
		{\bibfnamefont {A.~A.~P.}\ \bibnamefont {Trichet}}, \bibinfo {author}
		{\bibfnamefont {D.~M.}\ \bibnamefont {Whittaker}}, \bibinfo {author}
		{\bibfnamefont {P.~M.}\ \bibnamefont {Walker}}, \bibinfo {author}
		{\bibfnamefont {F.}~\bibnamefont {Fras}}, \bibinfo {author} {\bibfnamefont
			{E.}~\bibnamefont {Clarke}}, \bibinfo {author} {\bibfnamefont {J.~M.}\
			\bibnamefont {Smith}}, \bibinfo {author} {\bibfnamefont {M.~S.}\ \bibnamefont
			{Skolnick}},\ and\ \bibinfo {author} {\bibfnamefont {D.~N.}\ \bibnamefont
			{Krizhanovskii}},\ }\bibfield  {title} {\bibinfo {title} {Spin textures of
			exciton-polaritons in a tunable microcavity with large {TE-TM} splitting},\
	}\href {https://doi.org/10.1103/PhysRevLett.115.246401} {\bibfield  {journal}
		{\bibinfo  {journal} {Phys. Rev. Lett.}\ }\textbf {\bibinfo {volume} {115}},\
		\bibinfo {pages} {246401} (\bibinfo {year} {2015})}\BibitemShut {NoStop}%
	\bibitem [{\citenamefont {Leykam}\ and\ \citenamefont
		{Desyatnikov}(2016)}]{leykam.aipx.1.101.2016}%
	\BibitemOpen
	\bibfield  {author} {\bibinfo {author} {\bibfnamefont {D.}~\bibnamefont
			{Leykam}}\ and\ \bibinfo {author} {\bibfnamefont {A.~S.}\ \bibnamefont
			{Desyatnikov}},\ }\bibfield  {title} {\bibinfo {title} {Conical intersections
			for light and matter waves},\ }\href
	{https://doi.org/10.1080/23746149.2016.1144482} {\bibfield  {journal}
		{\bibinfo  {journal} {Adv. Phys. X}\ }\textbf {\bibinfo {volume} {1}},\
		\bibinfo {pages} {101} (\bibinfo {year} {2016})}\BibitemShut {NoStop}%
	\bibitem [{\citenamefont {Xiao}\ \emph {et~al.}(2010)\citenamefont {Xiao},
		\citenamefont {Chang},\ and\ \citenamefont {Niu}}]{xiao.rmp.82.1959.2010}%
	\BibitemOpen
	\bibfield  {author} {\bibinfo {author} {\bibfnamefont {D.}~\bibnamefont
			{Xiao}}, \bibinfo {author} {\bibfnamefont {M.-C.}\ \bibnamefont {Chang}},\
		and\ \bibinfo {author} {\bibfnamefont {Q.}~\bibnamefont {Niu}},\ }\bibfield
	{title} {\bibinfo {title} {Berry phase effects on electronic properties},\
	}\href {https://doi.org/10.1103/RevModPhys.82.1959} {\bibfield  {journal}
		{\bibinfo  {journal} {Rev. Mod. Phys.}\ }\textbf {\bibinfo {volume} {82}},\
		\bibinfo {pages} {1959} (\bibinfo {year} {2010})}\BibitemShut {NoStop}%
	\bibitem [{\citenamefont {Fuchs}\ \emph {et~al.}(2010)\citenamefont {Fuchs},
		\citenamefont {Pi\'echon}, \citenamefont {Goerbig},\ and\ \citenamefont
		{Montambaux}}]{fuchs.epjb.77.351.2010}%
	\BibitemOpen
	\bibfield  {author} {\bibinfo {author} {\bibfnamefont {J.~N.}\ \bibnamefont
			{Fuchs}}, \bibinfo {author} {\bibfnamefont {F.}~\bibnamefont {Pi\'echon}},
		\bibinfo {author} {\bibfnamefont {M.~O.}\ \bibnamefont {Goerbig}},\ and\
		\bibinfo {author} {\bibfnamefont {G.}~\bibnamefont {Montambaux}},\ }\bibfield
	{title} {\bibinfo {title} {Topological {Berry} phase and semiclassical
			quantization of cyclotron orbits for two dimensional electrons in coupled
			band models},\ }\href {https://doi.org/10.1140/epjb/e2010-00259-2} {\bibfield
		{journal} {\bibinfo  {journal} {Eur. Phys. J. B}\ }\textbf {\bibinfo
			{volume} {77}},\ \bibinfo {pages} {351} (\bibinfo {year} {2010})}\BibitemShut
	{NoStop}%
	\bibitem [{\citenamefont {Fukui}\ \emph {et~al.}(2005)\citenamefont {Fukui},
		\citenamefont {Hatsugai},\ and\ \citenamefont
		{Suzuki}}]{fukui.jpsj.74.1674.2005}%
	\BibitemOpen
	\bibfield  {author} {\bibinfo {author} {\bibfnamefont {T.}~\bibnamefont
			{Fukui}}, \bibinfo {author} {\bibfnamefont {Y.}~\bibnamefont {Hatsugai}},\
		and\ \bibinfo {author} {\bibfnamefont {H.}~\bibnamefont {Suzuki}},\
	}\bibfield  {title} {\bibinfo {title} {Chern numbers in discretized
			{Brillouin} zone: Efficient method of computing (spin) {Hall} conductances},\
	}\href {https://doi.org/10.1143/JPSJ.74.1674} {\bibfield  {journal} {\bibinfo
			{journal} {J. Phys. Soc. Jpn.}\ }\textbf {\bibinfo {volume} {74}},\ \bibinfo
		{pages} {1674} (\bibinfo {year} {2005})}\BibitemShut {NoStop}%
	\bibitem [{\citenamefont {Xie}\ \emph {et~al.}(2021)\citenamefont {Xie},
		\citenamefont {Wang}, \citenamefont {Zhang}, \citenamefont {Zhan},
		\citenamefont {Jiang}, \citenamefont {Lu},\ and\ \citenamefont
		{Chen}}]{xie.nrp.3.520.2021}%
	\BibitemOpen
	\bibfield  {author} {\bibinfo {author} {\bibfnamefont {B.}~\bibnamefont
			{Xie}}, \bibinfo {author} {\bibfnamefont {H.-X.}\ \bibnamefont {Wang}},
		\bibinfo {author} {\bibfnamefont {X.}~\bibnamefont {Zhang}}, \bibinfo
		{author} {\bibfnamefont {P.}~\bibnamefont {Zhan}}, \bibinfo {author}
		{\bibfnamefont {J.-H.}\ \bibnamefont {Jiang}}, \bibinfo {author}
		{\bibfnamefont {M.}~\bibnamefont {Lu}},\ and\ \bibinfo {author}
		{\bibfnamefont {Y.}~\bibnamefont {Chen}},\ }\bibfield  {title} {\bibinfo
		{title} {Higher-order band topology},\ }\href
	{https://doi.org/10.1038/s42254-021-00323-4} {\bibfield  {journal} {\bibinfo
			{journal} {Nat. Rev. Phys.}\ }\textbf {\bibinfo {volume} {3}},\ \bibinfo
		{pages} {520} (\bibinfo {year} {2021})}\BibitemShut {NoStop}%
	\bibitem [{\citenamefont {Lin}\ \emph {et~al.}(2023)\citenamefont {Lin},
		\citenamefont {Wang}, \citenamefont {Liu}, \citenamefont {Xue}, \citenamefont
		{Zhang}, \citenamefont {Chong},\ and\ \citenamefont
		{Jiang}}]{lin.nrp.5.483.2023}%
	\BibitemOpen
	\bibfield  {author} {\bibinfo {author} {\bibfnamefont {Z.-K.}\ \bibnamefont
			{Lin}}, \bibinfo {author} {\bibfnamefont {Q.}~\bibnamefont {Wang}}, \bibinfo
		{author} {\bibfnamefont {Y.}~\bibnamefont {Liu}}, \bibinfo {author}
		{\bibfnamefont {H.}~\bibnamefont {Xue}}, \bibinfo {author} {\bibfnamefont
			{B.}~\bibnamefont {Zhang}}, \bibinfo {author} {\bibfnamefont
			{Y.}~\bibnamefont {Chong}},\ and\ \bibinfo {author} {\bibfnamefont {J.-H.}\
			\bibnamefont {Jiang}},\ }\bibfield  {title} {\bibinfo {title} {Topological
			phenomena at defects in acoustic, photonic and solid-state lattices},\ }\href
	{https://doi.org/10.1038/s42254-023-00602-2} {\bibfield  {journal} {\bibinfo
			{journal} {Nat. Rev. Phys.}\ }\textbf {\bibinfo {volume} {5}},\ \bibinfo
		{pages} {483} (\bibinfo {year} {2023})}\BibitemShut {NoStop}%
	\bibitem [{\citenamefont {Ren}\ \emph {et~al.}(2023)\citenamefont {Ren},
		\citenamefont {Arkhipova}, \citenamefont {Zhang}, \citenamefont {Kartashov},
		\citenamefont {Wang}, \citenamefont {Zhuravitskii}, \citenamefont {Skryabin},
		\citenamefont {Dyakonov}, \citenamefont {Kalinkin}, \citenamefont {Kulik},
		\citenamefont {Kompanets}, \citenamefont {Chekalin},\ and\ \citenamefont
		{Zadkov}}]{ren.light.12.194.2023}%
	\BibitemOpen
	\bibfield  {author} {\bibinfo {author} {\bibfnamefont {B.}~\bibnamefont
			{Ren}}, \bibinfo {author} {\bibfnamefont {A.~A.}\ \bibnamefont {Arkhipova}},
		\bibinfo {author} {\bibfnamefont {Y.}~\bibnamefont {Zhang}}, \bibinfo
		{author} {\bibfnamefont {Y.~V.}\ \bibnamefont {Kartashov}}, \bibinfo {author}
		{\bibfnamefont {H.}~\bibnamefont {Wang}}, \bibinfo {author} {\bibfnamefont
			{S.~A.}\ \bibnamefont {Zhuravitskii}}, \bibinfo {author} {\bibfnamefont
			{N.~N.}\ \bibnamefont {Skryabin}}, \bibinfo {author} {\bibfnamefont {I.~V.}\
			\bibnamefont {Dyakonov}}, \bibinfo {author} {\bibfnamefont {A.~A.}\
			\bibnamefont {Kalinkin}}, \bibinfo {author} {\bibfnamefont {S.~P.}\
			\bibnamefont {Kulik}}, \bibinfo {author} {\bibfnamefont {V.~O.}\ \bibnamefont
			{Kompanets}}, \bibinfo {author} {\bibfnamefont {S.~V.}\ \bibnamefont
			{Chekalin}},\ and\ \bibinfo {author} {\bibfnamefont {V.~N.}\ \bibnamefont
			{Zadkov}},\ }\bibfield  {title} {\bibinfo {title} {Observation of nonlinear
			disclination states},\ }\href {https://doi.org/10.1038/s41377-023-01235-x}
	{\bibfield  {journal} {\bibinfo  {journal} {Light Sci. Appl.}\ }\textbf
		{\bibinfo {volume} {12}},\ \bibinfo {pages} {194} (\bibinfo {year}
		{2023})}\BibitemShut {NoStop}%
	\bibitem [{\citenamefont {Zhong}\ \emph
		{et~al.}(2024{\natexlab{b}})\citenamefont {Zhong}, \citenamefont {Kompanets},
		\citenamefont {Zhang}, \citenamefont {Kartashov}, \citenamefont {Cao},
		\citenamefont {Li}, \citenamefont {Zhuravitskii}, \citenamefont {Skryabin},
		\citenamefont {Dyakonov}, \citenamefont {Kalinkin}, \citenamefont {Kulik},
		\citenamefont {Chekalin},\ and\ \citenamefont
		{Zadkov}}]{zhong.light.13.264.2024}%
	\BibitemOpen
	\bibfield  {author} {\bibinfo {author} {\bibfnamefont {H.}~\bibnamefont
			{Zhong}}, \bibinfo {author} {\bibfnamefont {V.~O.}\ \bibnamefont
			{Kompanets}}, \bibinfo {author} {\bibfnamefont {Y.}~\bibnamefont {Zhang}},
		\bibinfo {author} {\bibfnamefont {Y.~V.}\ \bibnamefont {Kartashov}}, \bibinfo
		{author} {\bibfnamefont {M.}~\bibnamefont {Cao}}, \bibinfo {author}
		{\bibfnamefont {Y.}~\bibnamefont {Li}}, \bibinfo {author} {\bibfnamefont
			{S.~A.}\ \bibnamefont {Zhuravitskii}}, \bibinfo {author} {\bibfnamefont
			{N.~N.}\ \bibnamefont {Skryabin}}, \bibinfo {author} {\bibfnamefont {I.~V.}\
			\bibnamefont {Dyakonov}}, \bibinfo {author} {\bibfnamefont {A.~A.}\
			\bibnamefont {Kalinkin}}, \bibinfo {author} {\bibfnamefont {S.~P.}\
			\bibnamefont {Kulik}}, \bibinfo {author} {\bibfnamefont {S.~V.}\ \bibnamefont
			{Chekalin}},\ and\ \bibinfo {author} {\bibfnamefont {V.~N.}\ \bibnamefont
			{Zadkov}},\ }\bibfield  {title} {\bibinfo {title} {Observation of nonlinear
			fractal higher order topological insulator},\ }\href
	{https://doi.org/10.1038/s41377-024-01611-1} {\bibfield  {journal} {\bibinfo
			{journal} {Light Sci. Appl.}\ }\textbf {\bibinfo {volume} {13}},\ \bibinfo
		{pages} {264} (\bibinfo {year} {2024}{\natexlab{b}})}\BibitemShut {NoStop}%
	\bibitem [{\citenamefont {Kompanets}\ \emph {et~al.}(2025)\citenamefont
		{Kompanets}, \citenamefont {Feng}, \citenamefont {Zhang}, \citenamefont
		{Kartashov}, \citenamefont {Li}, \citenamefont {Zhuravitskii}, \citenamefont
		{Skryabin}, \citenamefont {Kireev}, \citenamefont {Dyakonov}, \citenamefont
		{Kalinkin}, \citenamefont {Shang}, \citenamefont {Kulik}, \citenamefont
		{Chekalin},\ and\ \citenamefont {Zadkov}}]{kompanets.am.37.2500556.2025}%
	\BibitemOpen
	\bibfield  {author} {\bibinfo {author} {\bibfnamefont {V.~O.}\ \bibnamefont
			{Kompanets}}, \bibinfo {author} {\bibfnamefont {S.}~\bibnamefont {Feng}},
		\bibinfo {author} {\bibfnamefont {Y.}~\bibnamefont {Zhang}}, \bibinfo
		{author} {\bibfnamefont {Y.~V.}\ \bibnamefont {Kartashov}}, \bibinfo {author}
		{\bibfnamefont {Y.}~\bibnamefont {Li}}, \bibinfo {author} {\bibfnamefont
			{S.~A.}\ \bibnamefont {Zhuravitskii}}, \bibinfo {author} {\bibfnamefont
			{N.~N.}\ \bibnamefont {Skryabin}}, \bibinfo {author} {\bibfnamefont {A.~V.}\
			\bibnamefont {Kireev}}, \bibinfo {author} {\bibfnamefont {I.~V.}\
			\bibnamefont {Dyakonov}}, \bibinfo {author} {\bibfnamefont {A.~A.}\
			\bibnamefont {Kalinkin}}, \bibinfo {author} {\bibfnamefont {C.}~\bibnamefont
			{Shang}}, \bibinfo {author} {\bibfnamefont {S.~P.}\ \bibnamefont {Kulik}},
		\bibinfo {author} {\bibfnamefont {S.~V.}\ \bibnamefont {Chekalin}},\ and\
		\bibinfo {author} {\bibfnamefont {V.~N.}\ \bibnamefont {Zadkov}},\ }\bibfield
	{title} {\bibinfo {title} {Observation of nonlinear topological corner
			states originating from different spectral charges},\ }\href
	{https://doi.org/10.1002/adma.202500556} {\bibfield  {journal} {\bibinfo
			{journal} {Adv. Mater.}\ }\textbf {\bibinfo {volume} {37}},\ \bibinfo {pages}
		{2500556} (\bibinfo {year} {2025})}\BibitemShut {NoStop}%
\end{thebibliography}

\end{document}